\documentclass[10pt,aps,prd,twocolumn,floatfix,shownopacs,preprintnumbers,nofootinbib,longbibliography]{revtex4-1}
\pdfoutput=1
\usepackage{amsmath,amsfonts,amsthm,bm,amssymb,mathtools,tabu}
\usepackage{graphicx}
\usepackage{gensymb}
\usepackage[caption=false]{subfig}
\usepackage{epstopdf}
\usepackage{csquotes}
\usepackage[T1]{fontenc}
\usepackage{lmodern}
\usepackage{mathtools}
\usepackage[colorlinks=true,breaklinks=true]{hyperref}
\hypersetup{allcolors=[rgb]{0.0 0.0 0.6},linkcolor=[rgb]{0.8 0.0 0.4}}   
\usepackage{slashed}             
\usepackage{url}
\usepackage[detect-none]{siunitx}
\usepackage{color}  
\usepackage{comment}              
\usepackage[dvipsnames]{xcolor}

\DeclareTextCommand{\textprime}{\encodingdefault}{%
  \mbox{$\m@th'\kern-\scriptspace$}%
}
\DeclarePairedDelimiter\abs{\lvert}{\rvert}%
\DeclarePairedDelimiter\norm{\lVert}{\rVert}%
\makeatletter
\let\oldabs\abs
\def\abs{\@ifstar{\oldabs}{\oldabs*}}
\let\oldnorm\norm
\def\norm{\@ifstar{\oldnorm}{\oldnorm*}}
\makeatother

\allowdisplaybreaks

\bibpunct{[}{]}{,}{n}{}{,}

\renewcommand{\vec}[1]{{\mathbf{#1}}}
\providecommand{\abs}[1]{\lvert#1\rvert}
\providecommand{\norm}[1]{\lVert#1\rVert}

\definecolor{blue}{rgb}{0.0,0.0,1.0}

\begin{document}
\title{Fast Neutrino Flavor Conversion at Late Time}
\author{Soumya Bhattacharyya}
\email{soumya.bhattacharyya@theory.tifr.res.in}
\affiliation{Tata Institute of Fundamental Research, Homi Bhabha
Road, Mumbai 400005, India}
\author{Basudeb Dasgupta}
\email{bdasgupta@theory.tifr.res.in}
\thanks{\scriptsize \!\!  \href{http://orcid.org/0000-0001-6714-0014}{orcid.org/0000-0001-6714-0014}}
\affiliation{Tata Institute of Fundamental Research, Homi Bhabha
Road, Mumbai 400005, India}
\date{May 25, 2020}

\begin{abstract} 
We study the fully nonlinear fast flavor evolution of neutrinos in 1+1 dimensions. Our numerical analysis shows that at late time the system reaches an approximately steady state. Using the steady state approximation we analytically show that the spatial variation of the polarization vectors is given by their precession around a common axis, which itself has a motion reminiscent of a gyroscopic pendulum. We then show that the steady state solution to the equations of motion cannot be separated in position and velocity, that is the motion is not collective in the usual sense. However, the fast evolution allows spectral-swap-like dynamics leading to partial decoherence over a range of velocities, constrained by conservation of lepton number(s). Finally, we numerically show that at late time  the transverse components of the polarization vectors become randomly oriented at different spatial locations for any velocity mode and lepton asymmetry.
\end{abstract}
\preprint{TIFR/TH/20-15}

\maketitle

\section{Introduction}
\label{sec:intro}
Neutrinos emitted by stars present valuable opportunities to study neutrino properties~\cite{Raffelt:1996wa}. While solar neutrinos have famously helped zero in on the large mixing angle scenario, neutrinos from supernovae may yet provide a unique opportunity to study neutrino-neutrino interactions -- a crucial piece of the standard model of particle physics that has not been tested directly.

The rate of neutrino oscillations is typically dictated by the vacuum oscillation frequency,  $\omega$, and the matter potential, $\lambda$~\cite{PhysRevD.17.2369, Mikheev:1987jp, Mikheev:1987qk}. Until the early 2000s, it was believed that this paradigm was sufficient to describe neutrino oscillations inside supernovae as well~\cite{Dighe:1999bi}. At that time, the outstanding problem of the field appeared to be to understand the effect of fluctuations in the background matter density~\cite{Dasgupta:2005wn, Fogli:2006xy, Friedland:2006ta}.

Following the pioneering papers by Pantaleone~\mbox{\cite{Pantaleone:1992xh, Pantaleone:1992eq}}, however, it became clear that the issue is more subtle~\cite{Kostelecky:1994dt, Pastor:2001iu}. Owing to the large neutrino density, even free-streaming neutrinos experience significant forward-scattering off other neutrinos. Such scattering leads to a self-interaction potential, $\mu \gg \omega$, that is proportional to the neutrino density and can dominate over the vacuum term. As a result, a gamut of new collective flavor transformations can occur inside supernovae.

The so-called ``slow'' collective effects, with an intrinsic rate $\sim\sqrt{\omega \mu}$, are already faster than usual neutrino oscillations. These lead to a variety of new phenomena, e.g., 
synchronization~\cite{Kostelecky:1994dt}, bipolar oscillations~\mbox{\cite{Pastor:2001iu, Duan:2005cp, Duan:2006an, Hannestad:2006nj, Johns_2018}}, spectral swaps~\cite{Raffelt:2007cb, Raffelt:2007xt, Fogli:2007bk, Dasgupta:2009mg}, three-flavor effects~\mbox{\cite{Dasgupta:2007ws, Duan:2008za, Dasgupta:2008cd, Doring:2019axc}}, multi-angle effects~\cite{EstebanPretel:2007bz, Duan:2010bf,  Chakraborty:2011gd, Saviano:2012yh},
decoherence~\cite{Raffelt:2007yz, EstebanPretel:2007ec, Hansen:2018apu, Hansen:2019iop}, and linear instabilities~\cite{Banerjee:2011fj}, including those that break symmetries of direction~\cite{Raffelt:2013isa, Raffelt:2013rqa}, space~\cite{Mangano:2014zda, Duan:2014gfa}, and time~\cite{Abbar:2015mca,  Dasgupta:2015iia}. Related developments, that followed the influential papers by Duan, Fuller, Carslon and Qian, and their phenomenological consequences have been reviewed in Refs.\,\cite{Duan:2010bg, Mirizzi:2015eza, Chakraborty:2016yeg, Horiuchi:2017sku}; see also references therein.

Ray Sawyer pointed out that much more rapid ``fast'' flavor conversions can take place~\cite{Sawyer:2005jk, Sawyer:2015dsa, Chakraborty:2016lct, Dasgupta:2016dbv}. These have a frequency $\sim\mu$, and might have a much more drastic effect for neutrino physics~\cite{Izaguirre:2016gsx, Capozzi:2017gqd, Das:2017iuj,  Dasgupta:2017oko, Dighe:2017sur, Abbar:2017pkh, Morinaga:2018aug, Capozzi:2018rzl, Dasgupta:2018ulw,  Capozzi:2018clo, Airen:2018nvp,  Abbar:2018beu, Yi:2019hrp, Martin:2019kgi, Capozzi:2019lso, Martin:2019gxb, Chakraborty:2019wxe, Johns:2019izj, Shalgar:2019qwg, Cherry:2019vkv, Abbar:2020fcl} as well as supernova astrophysics~\cite{Tamborra:2017ubu, Azari:2019jvr, Morinaga:2019wsv, Nagakura:2019sig, DelfanAzari:2019tez, Abbar:2019zoq, Glas:2019ijo}. The criterion for fast conversions to occur appears to be related to that for slow  conversions, i.e., the difference of neutrino and antineutrino flux distributions in the momentum space must have a zero crossing~\cite{Dasgupta:2009mg}, though a more detailed understanding still remains wanting.

The flavor evolution of a dense neutrino gas is governed by a large number of coupled nonlinear partial differential equations. These are almost always very difficult to solve. Although linearized stability analysis is useful to ascertain if and when fast conversion takes place, it cannot directly answer the  question -- what is the impact of fast flavor conversion on observable neutrino fluxes or the explosion mechanism? This is a significantly harder problem that requires understanding the nature of the solution in the nonlinear regime. A step in this direction was taken by Sen and one of the present authors~\cite{Dasgupta:2017oko}, where the flavor evolution of a 4-beam model in 0+1 dimension was understood in the fully nonlinear regime. 

In this work, we take another step in the same direction. We consider a dense neutrino gas in 1+1 dimensions, with a spectrum of velocity modes, and analyze the coupled flavor evolution of the neutrino system into the nonlinear regime. Our numerical analysis suggests that the system reaches an approximately steady state at late time. In the steady state approximation, we analytically show that the spatial variation of the polarization vectors is given by their precession around a gyrating flavor pendulum with a fixed length, spin, and energy, and the solution is not collective. The polarization vectors, when averaged over space, however, exhibit complete (partial) decoherence for zero (nonzero) lepton asymmetry. For partial decoherence, the non-vanishing range of velocity modes is dictated by conservation of lepton numbers. This kinematic decoherence stems from randomization of the transverse components. Numerical examples confirm these analytical insights.

\clearpage

The paper is structured as follows: Sec.\,\ref{ana} recollects the equations of motion, followed by our analytical results on the nature of the solution, conserved quantities, partial decoherence, and its dependence on lepton asymmetry. Sec.\,\ref{Num} has our numerical results and their comparison with analytical claims. We conclude in Sec.\,\ref{Con}.

\section{Analytical Results}\label{ana}
\subsection{Equations of Motion}
Neglecting momentum-changing collisions, the space and time evolution of two flavors of neutrinos with velocity ${\bf v}$ and vacuum oscillation frequency $\omega= {|\Delta m^{2}|/(2{\cal E})}$ is given by~\cite{Chakraborty:2016lct, Dasgupta:2016dbv},  
\begin{equation}\label{1}
\begin{split}
\left(\partial_{t}+\vec{v}.\nabla_{\bf x}\right)\vec{P}_{\omega, \textbf{v}}({\bf x}, t) = {\bf H}_{\omega, \textbf{v}}({\bf x},t)\times \vec{P}_{\omega, \textbf{v}}({\bf x}, t)\,.
\end{split}
\end{equation}
Here $\vec{P}_{\omega, \textbf{v}}= g_{\omega, \textbf{v}}\vec{S}_{\omega, \textbf{v}}= g_{\omega, \textbf{v}}\big(s_{v}^{(1)}, s_{v}^{(2)}, s_{v}^{(3)}\big)^{T}$ is a so-called polarization vector, that encapsulates the flavor composition of neutrino mode $\omega, {\bf v}$, and ${\bf H}_{\omega, \textbf{v}}={\bf H}^{\rm vac}_{\omega}+{\bf H}^{\rm mat}+{\bf H}^{\rm self}_{\omega,{\bf v}}$ is the Bloch vector representation of the flavor-evolution Hamiltonian for each mode. The vector ${\bf H}^{\rm vac}_{\omega} = \omega \vec{B}$ encodes vacuum oscillations with $\vec{B} = \big(\sin{2\theta}, 0, \cos{2\theta}\big)^{T}$, the ${\bf H}^{\rm mat} = \lambda \vec{L}$ encodes the matter effects with $\lambda = \sqrt{2} G_{F} (n_{e^-}-n_{e^+})$ and $\vec{L} = \big(0, 0,1\big)^{T}$, and ${\bf H}^{\rm self}_{\omega,{\bf v}} = \mu \int d\Gamma{'}\big(1-\textbf{v}.\textbf{v}{'}\big) \vec{P}_{\omega{'}, \textbf{v}{'}}(x, t)$ encodes the self-interactions, with $\mu$ being the $\nu-\nu$ potential. Note that we work in a basis $\{\hat{\bf e}_1, \hat{\bf e}_2, \hat{\bf e}_3\}$, where the Bloch vector for a $\nu_e$ points along the direction $\hat{\bf e}_3$.

In the fast flavor limit the vacuum and matter term in Eq.\eqref{1} are negligible compared to the neutrino potential term, so the solution for $\vec{P}_{\omega, \textbf{v}}$ becomes $\omega$-independent. The self-term then enters the Hamiltonian only through the electron lepton number (ELN) distribution, i.e., the  difference  of occupation number  densities  integrated  over  energy, defined by $G_{\textbf{v}} = \int d\omega \, g_{\omega, \textbf{v}}$. Thereby, allowing us to rewrite Eq.\eqref{1} in 1+1\,D as
\begin{equation}\label{2}
\big(\partial_{t}+v\partial_{x}\big)\vec{P}_{v}(x, t) = \int_{-1}^{1} dv{'}\left(1-vv{'}\right) \vec{P}_{{v}{'}}(x, t) \times \vec{P}_{v}(x, t)\,,
\end{equation}
where $\vec{P}_{v}(x, t) = G_{v}\vec{S}_{v}(x, t) = G_{v} \big(s_{v}^{(1)}, s_{v}^{(2)}, s_{v}^{(3)}\big)^{T}$. We choose our units in Eq.\eqref{2} such that neutrino self-interaction potential $\mu$ is 1, with length and time expressed in units of $\mu^{-1}$. $G_{v}$ encodes the amount of lepton asymmetry of the system as a function of velocity modes.

One can obtain some more insight into the equations of motion, by expanding $\vec{P}_{v}(x,t)$ for each velocity mode in terms of Legendre polynomials, $L_
n(v)$. Using the expansion $\vec{P}_{v}(x,t) = \sum_{n=0}^{\infty}\big(n+\frac{1}{2}\big) \vec{M}_{n}(x,t) L_{n}(v)$, and using the orthogonal property of Legendre polynomials, i.e., $\int_{-1}^{1} L_{r}(v)L_{n}(v)dv = 2\,\delta_{rn}/(2r+1)$, following Ref.\,\cite{Raffelt:2007yz} one can rewrite Eq.\,\eqref{2} as
\begin{equation}\label{3a}
\begin{split}
(\partial_t+v\partial_{x})\vec{P}_{v}(x,t) = \big({\vec{M}_{0}(x,t)}-v\vec{M}_{1}(x,t)\big) \times \vec{P}_{v}(x,t)\,.
\end{split}
\end{equation}

\subsection{Steady State}

We conjecture that at late time  the system becomes approximately stationary in time, which says that in Eq.\eqref{2} we can drop the $t$-dependence from all the quantities. We will verify this conjecture in our numerical survey described in Sec.\,\ref{Num}.

In the steady state one can write Eq.\eqref{2} as
\begin{equation}\label{3}
\begin{split}
d_{x}\vec{P}_{v}(x) = \bigg(\frac{\vec{M}_{0}(x)}{v}-\vec{M}_{1}(x)\bigg) \times \vec{P}_{v}(x)\,,
\end{split}
\end{equation}
and the equation for each multipole moment $\vec{M}_{r}(x)$ as
\begin{equation}\label{4}
{d_{x}}\vec{M}_{r}(x)=\vec{M}_{0}(x) \times \sum_{n=0}^{\infty} \ell_{rn} \vec{M}_{n}(x)- \vec{M}_{1}(x) \times \vec{M}_{r}(x)\,, 
\end{equation}
where
\begin{equation}\label{5}
\ell_{rn} = \left(n+\frac{1}{2}\right)\int_{-1}^{1} \frac{L_{n}(v)L_{r}(v)}{v} dv\,.
\end{equation}
From Eq.\eqref{4}, for $r = 1$ and using $\ell_{1n} = (2n+1)\delta_{0n}$, one gets that $\vec{M}_{1}(x)$ is constant in space. This means that all the polarization vectors of different velocity modes precess about a fixed axis $\vec{M}_{1}$ with the same frequency. This common motion can be hidden away by considering a rotating frame, using the transformation
\begin{equation}\label{6}
 \widetilde{d}_{x}\vec{\widetilde{P}}_{v}(x) = {d_{x}}\vec{P}_{v}(x)+\vec{M}_{1}\times \vec{P}_{v}(x)\,,
\end{equation}
where we introduce the notation that all the quantities in the rotating frame are denoted by tilde, spatial derivatives in the rotating frame as $\widetilde{d}_{x}$, etc. 

Using the rotation formula described in Eq.\eqref{6}, the equations of motion for the system at late time,  described by Eqs.\eqref{3} and \eqref{4} in the rotating frame, look like
\begin{subequations}
\begin{align}
\widetilde{d}_{x}\widetilde{\vec{P}}_{v}(x) = \frac{ \widetilde{\vec{M}}_{0}(x)}{v} \times \widetilde{\vec{P}}_{v}(x)\,, \label{7a}\\
\widetilde{d}_{x}\widetilde{\vec{M}}_{r}(x)=\widetilde{\vec{M}}_{0}(x) \times \sum_{n=0}^{\infty}\ell_{rn} \widetilde{\vec{M}}_{n}(x)\,. \label{7b}
\end{align}
\end{subequations}
Considering $r = 0$ and summing both sides of Eq.$\eqref{7b}$ with weightage factor of $\ell_{0r}$, from $r = 0$ to $r = \infty$, one can rewrite Eq.$\eqref{7b}$ as two coupled equations:
 \begin{subequations}
 \begin{align}
 \widetilde{d}_{x} \vec{\widetilde{M}}_{0}(x) =  \vec{\widetilde{D}}(x) \times \vec{\widetilde{M}}_{0}(x)\,, \label{8a}\\
 \widetilde{d}_{x} \vec{\widetilde{D}}(x) =  \vec{\widetilde{B}}(x)\times \vec{\widetilde{M}}_{0}(x)\,, \label{8b}
 \end{align}
 \end{subequations}
where,
\vspace{-0.5cm}
\begin{align}
 \vec{\widetilde{D}(x)} = -\sum_{n=0}^{\infty} \ell_{0n}\vec{\widetilde{M}}_{n}(x)\,,\label{9}\\[-0.5em]
 \vec{\widetilde{B}}(x) = \sum_{r, n=0}^{\infty} \ell_{0r}\ell_{rn}\vec{\widetilde{M}}_{n}(x)\,.\label{10}
\end{align}
Similarly summing both sides of Eq.$\eqref{7b}$, with a weightage factor of $\sum_{p = 0}^{\infty} \ell_{0p} \ell_{pr}$, from $r= 0$ to $r = \infty$, one obtains an equation for $\widetilde{\vec{B}}(x)$:
 \begin{equation}\label{11}
 \widetilde{d}_{x}\widetilde{\vec{B}}(x) =  \sum_{p, r, n=0}^{\infty}  \ell_{0p} \ell_{pr}\ell_{rn}  \bigg(\vec{\widetilde{M}}_{0}(x)\times \vec{\widetilde{M}}_{n}(x)\bigg)\,.
 \end{equation}
Eqs.\eqref{8a} and \eqref{8b} are nominally the same set of coupled equations which describe the ``gyroscopic pendulum'' (see Ref.~\cite{Hannestad:2006nj}), with the difference that $\widetilde{\vec{B}}(x)$, instead of being a constant, has an equation of motion in space described by Eq.\eqref{11}. Nevertheless, one can follow Ref.~\cite{Hannestad:2006nj} to derive an equation for a gyroscopic pendulum with a position-independent length, spin, and energy. Eq.\eqref{8a} clearly indicates $M_{0}$, i.e., the length of ${\widetilde{\vec{M}}_{0}(x)}$ which is the same in rotating and non-rotating frames, is a constant in space. Taking dot products with $\widetilde{\vec{D}}(x)$ or $\widetilde{\vec{B}}(x)$, respectively, on both sides of Eq.\eqref{8a}, and with $\widetilde{\vec{M}}_{0}(x)$ or $\widetilde{\vec{D}}(x)$, respectively, on both sides of Eq.\eqref{8b}, and with $\widetilde{\vec{M}}_{0}(x)$ for Eq.\eqref{11}, one gets the following set of equations:
 \begin{subequations}
 \begin{align}
 \widetilde{\vec{D}}(x)\cdot\widetilde{d}_{x}\widetilde{\vec{M}}_{0}(x) = 0\,,\label{12a}\\ 
 \widetilde{\vec{B}}(x)\cdot\widetilde{d}_{x}\widetilde{\vec{M}}_{0}(x) = \widetilde{\vec{B}}(x)\cdot\bigg(\widetilde{\vec{D}}(x) \times \widetilde{\vec{M}}_{0}(x)\bigg)\,, \label{12b}\\
 \widetilde{\vec{M}}_{0}(x)\cdot\widetilde{d}_{x}\widetilde{\vec{D}}(x) = 0\,, \label{12c}\\
 \widetilde{\vec{D}}(x)\cdot\widetilde{d}_{x}\widetilde{\vec{D}}(x) = \widetilde{\vec{D}}(x)\cdot\bigg(\widetilde{\vec{B}}(x) \times \widetilde{\vec{M}}_{0}(x)\bigg)\,, \label{12d}\\ 
 \widetilde{\vec{M}}_{0}(x)\cdot\widetilde{d}_{x}\widetilde{\vec{B}}(x) = 0\,. \label{12e}
 \end{align}
 \end{subequations}
Adding Eqs.\eqref{12a} and \eqref{12c} one obtains
\begin{equation}\label{13}
\widetilde{d}_{x}\bigg(\widetilde{\vec{M}}_{0}(x).\widetilde{\vec{D}}(x)\bigg) = 0\,,
\end{equation}
which implies that the ``spin''
\begin{equation}\label{14}
\widetilde{\vec{m}}_{0}(x)\cdot\widetilde{\vec{D}}(x) = \sigma = \rm{constant}\,,
\end{equation}
i.e., position-independent and owing to the steady-state approximation also time-independent. Here, $\widetilde{\vec{m}}_{0}(x) = {\widetilde{\vec{M}}_{0}(x)}/{M}_{0}$ is the unit vector along $\widetilde{\vec{M}}_{0}(x)$. Similarly addition of Eqs.\eqref{12b}, \eqref{12d}, and \eqref{12e} reveals energy conservation:
\begin{equation}\label{15}
\widetilde{d}_{x}\bigg(\widetilde{\vec{B}}(x)\cdot\widetilde{\vec{M}}_{0}(x)+\frac{1}{2}\widetilde{\vec{D}}(x)\cdot\widetilde{\vec{D}}(x)\bigg) = 0\,,
\end{equation}
which implies 
\begin{equation}\label{16}
\widetilde{\vec{B}}(x)\cdot\widetilde{\vec{M}}_{0}(x)+\frac{1}{2}\widetilde{\vec{D}}(x)\cdot\widetilde{\vec{D}}(x) = E = \rm{constant}\,.
\end{equation}
Taking a cross product of Eq.$\eqref{8a}$ with $\widetilde{\vec{M}}_{0}(x)$ one gets
\begin{equation}\label{17}
\begin{split}
{\widetilde{\vec{M}}_{0}(x) \times \widetilde{d}_{x}  \widetilde{\vec{M}}_{0}(x) = \hspace{4cm}}  \\
{-\Big(\widetilde{\vec{M}}_{0}(x)\cdot\widetilde{\vec{D}}(x)\Big) \widetilde{\vec{M}}_{0}(x)+ M_{0}^{2}\, \widetilde{\vec{D}}(x)\,.}
\end{split}
\end{equation}
Dividing both sides of Eq.\eqref{17} with $M_{0}^{2}$ and then using Eq.\eqref{14}, one can rewrite Eq.\eqref{17} as
\begin{equation}\label{18}
\widetilde{\vec{D}}(x) =  \widetilde{\vec{m}}_{0}(x) \times \widetilde{d}_{x} \widetilde{\vec{m}}_{0}(x)+\sigma\,\widetilde{\vec{m}}_{0}(x)\,.
\end{equation}
Differentiating Eq.\eqref{18} once, and using the spatial conservation of $\sigma$ along with Eq.\eqref{8b}, gives
\begin{equation}\label{19}
\begin{aligned}
\widetilde{\vec{m}}_{0}(x) \times \widetilde{d}^{\,2}_{x} \widetilde{\vec{m}}_{0}(x)+\sigma\,\widetilde{d}_{x} \widetilde{\vec{m}}_{0}(x) =  M_{0}  \widetilde{\vec{B}}(x)\times \widetilde{\vec{m}}_{0}(x)\,.
\end{aligned}
\end{equation}

The vector $\widetilde{\vec{M}}_{0}(x)$ in flavor space plays the role of a gyroscopic pendulum. It has a fixed length at all spatial locations, so that it is restricted to move on a sphere of a fixed radius $M_{0}$. According to Eq.\eqref{16} the energy of the pendulum is spatially invariant at late time, where the first term $\widetilde{\vec{B}}(x)\cdot\widetilde{\vec{M}}_{0}(x)$ is equivalent to the potential energy of $\widetilde{\vec{M}}_{0}(x)$ in an inhomogeneous magnetic field $\widetilde{\vec{B}}(x)$, and $\frac{1}{2}\widetilde{\vec{D}}(x)\cdot\widetilde{\vec{D}}(x)$ is the rotational energy of the system, with $\widetilde{\vec{D}}(x)$ playing the role of the orbital angular momentum of the system. Eq.\eqref{14} describes one more conserved quantity $\sigma$, which says that the component of the angular momentum $\widetilde{\vec{D}}(x)$, that is parallel to $\widetilde{\vec{M}}_{0}(x)$, is equivalent to the pendulum's spin and is spatially constant. 

The above analysis shows that at late time  the spatial structure of the solution is very simple: every $\widetilde{\vec{P}}_{v}$ has a precession about an axis $\widetilde{\vec{m}}_{0}(x)$ with a frequency $\frac{M_{0}}{v}$, where $\widetilde{\vec{m}}_{0}(x)$ itself has a motion equivalent to that of a gyroscopic pendulum in an inhomogeneous magnetic field, keeping a fixed length, spin and energy. This similarity to the gyroscopic pendulum solution, in a co-rotating frame, is however limited to the conserved quantities and a formal similarity of the equations of motion. The actual motion is different, because $\widetilde{\bf B}$ in this case has a nontrivial motion that is not necessarily much slower than that of the $\widetilde{\bf M}_n$. These conserved quantities are formally identical to those in Ref.\,\cite{Johns:2019izj}, but they have a slightly different interpretation. Unlike in the above, where exactly 0+1D or 1+0D was considered, ours are obtained under a steady state approximation in 1+1D and need not be \emph{exactly} conserved.

\subsubsection{Nature of the Solution}
We now prove that the solution at late time, described by Eq.\eqref{7a}, cannot be separable in position and velocity coordinates, i.e., it cannot be written in the form:
\begin{equation}\label{20}
\widetilde{\vec{P}}_{v} = G_{v} \widetilde{\vec{S}}_{v} =
 G_{v}\begin{pmatrix}
 f_{1}(x) h_{1}(v)\\ 
 f_{2}(x) h_{2}(v)\\
 f_{3}(x) h_{3}(v)                                                                                                                                                                            \end{pmatrix}\,.
\end{equation}
In our notation, the components of $\vec{S}_{v}$ in the rotated frame look like $\widetilde{\vec{S}}_{v} = \big(\widetilde{s}_{v}^{(1)}, \widetilde{s}_{v}^{(2)}, \widetilde{s}_{v}^{(3)}\big)^{T}$, with $\abs{\widetilde{\vec{S}}_{v}}=1$. For the above type of solution, clearly $f_{1}(x) = f_{2}(x) = f_{3}(x)$ is not possible. Otherwise, the normalization of $\widetilde{\vec{S}}_{v}$ for a fixed velocity mode will be different at different points in space, which is obviously unphysical. By a similar argument, $h_{1}(v) = h_{2}(v) = h_{3}(v)$ cannot be possible either, as that will lead to $\widetilde{\vec{S}}_{v}$ having different normalizations for different velocity modes at a fixed point in space.
Plugging Eq.\eqref{20} into Eq.\eqref{7a}, we get two separate sets of equations -- one governing the spatial dependence and the other with the velocity dependence of the full solution. The equations governing the velocity dependence look like
\begin{subequations}
	\begin{align}
	v h_{1}(v) = H_{2}h_{3}(v)-h_{2}(v)H_{3}\,,\label{21a}\\
	v h_{2}(v) = H_{3}h_{1}(v)-h_{3}(v)H_{1}\,,\label{21b}\\
	v h_{3}(v)  = H_{1}h_{2}(v)-h_{1}(v)H_{2}\,,\label{21c}
	\end{align}
\end{subequations}
where,
\begin{subequations}
	\begin{align}
	H_{1} = \int_{-1}^{1}G_{v} h_{1}(v)\,dv\,,\label{22a}\\
	H_{2} = \int_{-1}^{1}G_{v} h_{2}(v)\,dv\,,\label{22b}\\
	H_{3} = \int_{-1}^{1}G_{v} h_{3}(v)\,dv\,.\label{22c}
	\end{align}
\end{subequations}
For all modes with $v \neq 0$, Eqs.\eqref{21a}--\eqref{21c} can be satisfied only if $h_{1}(v)  = h_{2}(v) = h_{3}(v) = 0$. This already suggests that, again, no meaningful solutions exist. However, we should confirm that the spatial solutions do not diverge. The equations for the spatial dependence look like
\begin{subequations}
 \begin{align}
  \frac{d}{dx}f_{1}(x) = f_{2}(x)f_{3}(x)\,, \label{23a}\\
  \frac{d}{dx}f_{2}(x) = f_{1}(x)f_{3}(x)\,, \label{23b}\\
  \frac{d}{dx}f_{3}(x) = f_{1}(x)f_{2}(x)\,. \label{23c}
 \end{align}
\end{subequations}
To solve the above set of equations, we multiply Eqs.\eqref{23a}, \eqref{23b}, and \eqref{23c} by $f_{1}(x)$, $f_{2}(x)$, and $f_{3}(x)$, respectively, which gives 
\begin{equation}\label{24}
 \frac{d}{dx}\frac{f_{1}(x)^{2}}{2} =  \frac{d}{dx}\frac{f_{2}(x)^{2}}{2} =  \frac{d}{dx}\frac{f_{3}(x)^{2}}{2} = f_{1}(x) f_{2}(x) f_{3}(x)\,.
\end{equation}
Eq.\eqref{24} implies
\begin{subequations}
\begin{align}
 f_{1}(x)^{2} = f_{3}(x)^{2}+C_{1}\,,\label{25a} \\
 f_{2}(x)^{2} = f_{3}(x)^{2}+C_{2}\,,\label{25b}
 \end{align}
\end{subequations}
where $C_{1}$ and $C_{2}$ are integration constants. The choices $C_{1} = C_{2} = 0$ are disallowed, as that will mimic the case $f_{1}(x) = f_{2}(x) = f_{3}(x)$. Eqs.\eqref{25a}, \eqref{25b}, along with Eq.\eqref{23c}, give
 \begin{equation}\label{26}
  \frac{d}{dx}f_{3}(x) = \pm \sqrt{f_{3}(x)^{2}+C_{1}}\sqrt{f_{3}(x)^{2}+C_{2}}\,.
 \end{equation}
Solving Eq.\eqref{26} one gets
\begin{equation}\label{27}
 \frac{-i}{\sqrt{C_{2}}}F\bigg(\sin^{-1}\bigg[\sqrt{-1/C_{1}} f_{3}(x)\bigg], \frac{C_{1}}{C_{2}}\bigg) = x + C_{3}\,,
\end{equation}
where $F\bigg(\sin^{-1}\bigg[\sqrt{-1/C_{1}} f_{3}(x)\bigg], \frac{C_{1}}{C_{2}}\bigg)$ is an elliptic integral of first kind defined as
\begin{equation}\label{28}
 F(\phi, m) = \int_{0}^{\phi}(1-m\sin^{2}\theta)^{-1/2} d\theta\,,
\end{equation}
and $C_{3}$ being an integration constant. One has to basically invert Eq.\eqref{27} to get the behavior of $f_{3}(x)$ as a function of $x$.  This function is not infinite everywhere and concludes our analytical proof that there are no separable solutions in steady state. 

Still, this may be opaque, and to give a flavor for the solution, consider a special case with $C_{1} =  C_{2}$, and the solution for Eq.\eqref{26} then becomes
\begin{equation}\label{29}
f_{3}(x) = \pm C_{1}\tanh\bigg(\sqrt{C}_{1}x \pm \sqrt{C}_{1}C_{3}\bigg)\,.
\end{equation}
For any value of $C_{1}$ and $C_{3}$, the solutions for $f_{1}(x)$, $f_{2}(x)$, and $f_{3}(x)$, are finite -- with either oscillatory or constant behavior over all space. This implies that, for all the modes with $v \neq 0$ the solution for all the components of $\widetilde{\vec{S}}_{v}(x)$ will be exactly zero, giving rise to an unphysical solution. Another qualitative way to understand this is from Eq.\eqref{3}, where one can see that due to the spatial dependence of the velocity dependent term, $\frac{\vec{M}_{0}(x)}{v} \times \vec{P}_{v}(x)$, in the right hand side of the equation, it can not be rotated away by a rotation of $\vec{P}_{v}(x)$ in the position space. So, the equation of motion governing the spatial behavior for each $\vec{P}_{v}(x)$ cannot be same for every velocity mode $v$, indicating a non-separable solution in $x$ and $v$.

\subsection{Approach to Steady State}
In this section, we consider the approach to steady state, \emph{without} making the steady state approximation. Rather, we investigate how the spatially averaged polarization vectors rearrange themselves at late time, conserving lepton number and giving rise to flavor depolarization over a range of velocities. 

\subsubsection{Dependence on Lepton Asymmetry}
Polarization vectors cannot be completely depolarized if the lepton asymmetry is non-vanishing. To see this, we integrate both sides of  Eq.\eqref{2} over all velocity modes to get
\begin{equation}\label{30}
 \partial_{x}\vec{M}_{1}(x, t) + \partial_{t}\vec{M}_{0}(x, t)= 
 0\,.
\end{equation}
Performing a spatial average on both sides of Eq.\eqref{30} over the entire box with the periodic boundary condition, $\vec{M}_{1}(L, t) = \vec{M}_{1}(0, t)$, one gets
\begin{equation}\label{31}
 {d_{t}}\bigl\langle \vec{M}_{0}(t) \bigr\rangle = 0\,.
\end{equation}
This implies
\begin{equation}\label{32}
\bigl\langle \vec{M}_{0}( t)  \bigr\rangle   = \rm{constant}\,,
\end{equation}
where $\bigl\langle \vec{M}_{0}(t) \bigr\rangle$ is understood as
\begin{equation}\label{33}
\bigl\langle \vec{M}_{0}(t) \bigr\rangle = \frac{1}{L}\int_{0}^{L} \,dx\,\vec{M}_{0}(x, t)\,. 
\end{equation}
Using the approximate stationarity at late time, one can argue, from Eq.\eqref{30}, that $\vec{M}_{1}(x, t)$ is approximately constant all over space. The constant in Eq.\eqref{32} can be determined from the initial condition of the system, where all neutrinos are emitted as approximately flavor pure states from every point in space, i.e., $s_{v}^{(3)}(x, 0) = 1$ for all $x$ and $v$. One can write $\bigl\langle \vec{M}_{0}(t) \bigr\rangle$ as
\begin{equation}\label{34}
\begin{aligned}
 \bigl\langle \vec{M}_{0}(t) \bigr\rangle &= \begin{pmatrix}
\int_{-1}^{1} dv\,G_{v} \bigl\langle s_{v}^{(1)}(t) \bigr\rangle \\
\int_{-1}^{1} dv\,G_{v} \bigl\langle s_{v}^{(2)}(t) \bigr\rangle \\
\int_{-1}^{1} dv\,G_{v} \bigl\langle s_{v}^{(3)}(t) \bigr\rangle                                                                                                                                                                                                                                                \end{pmatrix}\,,
\end{aligned}
\end{equation}
which remains true at all times. The above initial conditions, and defining $\int_{-1}^{1}G_{v}dv = A$, gives rise to a set of three conditions that have to be satisfied even in the nonlinear regime:
\begin{subequations}
\begin{align}
 \int_{-1}^{1}dv\,G_{v} \bigl\langle s_{v}^{(1)} \bigr\rangle = \int_{-1}^{1}dv\,G_{v} \bigl\langle s_{v}^{(2)} \bigr\rangle = 0\,, \label{39a}\\
 \int_{-1}^{1}dv\,G_{v} \bigl\langle s_{v}^{(3)} \bigr\rangle = A\,. \label{39b}
 \end{align}
\end{subequations}
One way that Eqs.\eqref{39a} and \eqref{39b} can be satisfied at late time for systems with zero lepton asymmetry, i.e., $A = 0$, is if $\langle s_{v}^{(1)} \rangle = \langle s_{v}^{(2)} \rangle = \langle s_{v}^{(3)} \rangle = 0$ for every velocity mode $v$. This indicates a depolarization of the system. However, for systems with $A \neq 0$ such a simple solution can exist if $\langle s_{v}^{(3)} \rangle = 1$ for every $v$. This is possible only for an inert system where fast flavor conversion do not occur. Therefore, fast flavor conversion in the nonlinear regime can happen for $A \neq 0$, if different velocity modes $\langle s_{v}^{(3)} \rangle $ behave in such a velocity-dependent way that it preserves the value of the lepton asymmetry $A$. We will find, numerically, in case of $A$ being $+ve$ (or $-ve$),  $\bigl\langle s_{v}^{(3)} \bigr\rangle$ will remain much closer to $1$ for modes for which $G_{v}$ is $+ve$ (or $-ve$).

\subsubsection{Fast Depolarization}
Now we will show that a set of polarization vectors can undergo a ``fast depolarization'' -- i.e., the system exhibits a ``fast spectral split'' like evolution, similar to what appears in the foundational study by Raffelt and Smirnov\,\cite{Raffelt:2007cb, Raffelt:2007xt}, that leads to $s^{(3)}_{v}(x,t)$ flipping (and approximately vanishing) over a range of velocities, but constrained by conserved lepton numbers.

Consider Eq.\,\eqref{3a}. After averaging over space, we get
\begin{align}
\partial_t\langle{\bf P}_v(t)\rangle =\bigl\langle {\bf M}_0(t)\times{\bf P}_v(t)\bigr\rangle- v\bigl\langle {\bf M}_1(t)\times{\bf P}_v(t)\bigr\rangle\,.
\end{align}
The averaging is over the cross-product of the vectors but, because the polarization vectors vary very fast over space, one expects that  the averaging factorizes over the cross product, which gives
\begin{align}
\partial_t \langle{\bf P}_v(t)\rangle =& \bigl(\bigl\langle {\bf M}_0(t)\rangle- v\bigl\langle {\bf M}_1(t)\bigr\rangle\bigr)\times\langle{\bf P}_v(t)\rangle\,.
\label{eq:prec}
\end{align}
Integrating the above equation over all velocities, one recovers that $\langle {\bf M}_0\rangle$ is time-independent. However, multiplying the above equation with $v$ and then integrating over $v$ gives
\begin{align}
\partial_t\langle{\bf M}_1(t)\rangle =\bigl(\bigl\langle {\bf M}_0(t)\bigr\rangle+\bigl\langle {\bf X}(t)\bigr\rangle\bigr)\times\bigl\langle{\bf M}_1(t)\bigr\rangle\,,
\label{eq:m1prec}
\end{align}
where we define a new vector,
\begin{align}
\bigl\langle {\bf X}(t)\bigr\rangle = \int_{-1}^{+1}dv\,v^2 \bigl\langle {\bf P}(t)\bigr\rangle\,.
\end{align}
Eq.\eqref{eq:prec} ensures that each $\langle{\bf P}_v(t)\rangle$ precesses around its Hamiltonian,
\begin{align}
\bigl\langle{\bf H}_v(t)\bigr\rangle=\bigl\langle {\bf M}_0(t)\bigr\rangle- v\bigl\langle {\bf M}_1(t)\bigr\rangle\,.
\end{align}
Initially all polarization vectors are either aligned or anti-aligned to $\hat{\bf e}_3$, and over time $\bigl\langle {\bf M}_1(t)\bigr\rangle$  has dynamics, while the polarization vectors are dragged by $\bigl\langle{\bf H}_v(t)\bigr\rangle$. 

We assume that the polarization vectors stay close to their Hamiltonians, and thus remain in the plane formed by $\bigl\langle {\bf M}_0(t)\bigr\rangle$ and $\bigl\langle {\bf M}_1(t)\bigr\rangle$, ignoring the precession around their Hamiltonian. The vector $\bigl\langle {\bf X}(t)\bigr\rangle$ can be decomposed as
\begin{align}
\bigl\langle {\bf X}(t)\bigr\rangle=\alpha(t)\bigl\langle {\bf M}_0(t)\bigr\rangle +\bigl\langle {\bf X}_\perp(t)\bigr\rangle\,,
\end{align}
which, upon insertion in Eq.\eqref{eq:m1prec}, gives
\begin{equation}
\partial_t\langle{\bf M}_1(t)\rangle =\bigg(\big(1+\alpha(t)\big)\bigl\langle {\bf M}_0(t)\bigr\rangle + \bigl\langle {\bf X}_\perp(t)\bigr\rangle\bigg)\times\bigl\langle{\bf M}_1(t)\bigr\rangle\,.
\end{equation}
Note that $\bigl\langle {\bf X}(t)\bigr\rangle$ typically has dynamics at the same frequency as $\bigl\langle {\bf M}_1(t)\bigr\rangle$. So $\bigl\langle{\bf M}_1(t)\bigr\rangle$, that has the dynamics of a gyroscopic pendulum, can not only precess around $\bigl\langle {\bf M}_0(t)\bigr\rangle$ but can also have bipolar nutations.

In a frame that co-rotates with the plane formed by  $\bigl\langle {\bf M}_0(t)\bigr\rangle$ and $\bigl\langle {\bf M}_1(t)\bigr\rangle$, the Hamiltonian shifts by \mbox{$-(1+\alpha(t))\bigl\langle {\bf M}_0(t)\bigr\rangle$} due to pure precession of $\bigl\langle {\bf M}_1(t)\bigr\rangle$. Focussing on the $\hat{\bf e}_3$ component of the Hamiltonian, we find that
\begin{align}
\bigl\langle\widetilde{\bf H}_v^{(3)}(t)\bigr\rangle=-\alpha(t)\bigl\langle {\bf M}_0^{(3)}(t)\bigr\rangle- v\bigl\langle {\bf M}_1^{(3)}(t)\bigr\rangle\,.
\end{align}
The components of $\bigl\langle {\bf M}_0(t)\bigr\rangle$ and $\bigl\langle {\bf M}_1(t)\bigr\rangle$  along $\hat{\bf e}_3$ remain conserved if $\bigl\langle {\bf M}_1(t)\bigr\rangle$ has no nutation. In any case, in addition to Eq.\eqref{39b}, we can define another useful lepton number,
\begin{align}
\bigl\langle {\bf M}_1^{(3)}(t)\bigr\rangle=\int_{-1}^{+1} dv\,v\,G_v\bigl\langle s_{v}^{(3)} \bigr\rangle = B(t)\,.
\label{eq:Bcons}
\end{align}
The above stays constant while $\bigl\langle {\bf M}_1(t)\bigr\rangle$ precesses, but may flips to a different value if a bipolar instability is triggered instead.
The $\hat{\bf e}_3$ component of the co-rotating Hamiltonian is therefore,
\begin{align}
\bigl\langle\widetilde{\bf H}_v^{(3)}(t)\bigr\rangle=-\alpha(t)A-vB(t)\,.
\end{align}
Depending on how $\alpha(t)$ and $B(t)$ vary with time, the above can change its sign between some initial time and final time. 
The condition for such a sign-flip to occur, is
\begin{align}
\bigl(\alpha_{\rm ini}A+vB_{\rm ini}\bigr)\bigl(\alpha_{\rm fin}A+vB_{\rm fin}\bigr)<0\,,
\label{eq:swapcond}
\end{align}
where $(\ldots)_{\rm ini,\,fin}$ are at the beginning and end of the evolution. If the Hamiltonian for a velocity mode $v$ can change sign by fulfilling Eq.\eqref{eq:swapcond}, and obeying the constraints on $A$ and $B$, then that velocity mode following its Hamiltonian may flip its sign as well.

So far, our discussion closely follows the discussion in Refs.\,\cite{Raffelt:2007cb, Raffelt:2007xt}. There are a few subtle differences. Unlike in the case of bipolar swaps where only the lepton number $A$ needs to remain conserved, here one has two constraint equations, i.e., conservation of $A$ and conservation or flip of $B$, depending on whether nutations occur. Further, this derivation was for spatially averaged polarization vectors, $\langle{\bf P}_v(t)\rangle$, that do not necessarily maintain the same length as their unaveraged counterparts ${\bf P}_v(t)$. This point is quite crucial, as the astute reader will notice, otherwise there is no obvious source of irreversibility: unlike in the bipolar spectral swap where that is provided by a decreasing $\mu$, the irreversibility is provided by the relative dephasing, i.e., kinematic decoherence, of the polarization vectors, which is responsible for the irreversibility of $\alpha(t)$. Further, with simple choices of $G_v$, the swapping function of the $\bigl\langle s_{v}^{(3)} \bigr\rangle$ cannot be a mere sign-flip over a block of velocities, in general. Such a sign-change across a crossing can preserve $A$, but unless this block of $G_v$ is antisymmetric in $v$ such a flip doesn't obey the constraint on $B$.

More detailed exploration of fast depolarization will be published separately, but here we note a main qualitative feature: over time the $\bigl\langle s_{v}^{(3)} \bigr\rangle$ for a range of velocities become close to zero, even flipping their sign, but in the remaining range stay close to their initial state $\bigl\langle s_{v}^{(3)} \bigr\rangle=1$. Consider a simple case, where $A>0$, $B>0$ and $G_v$ has a single crossing at zero. It is possible that the modes with $v\approx+1$ do not flip sign or become small, whereas those with smaller $v$ either become very small or flip their sign (with change in magnitude, as well).  The converse however is not possible because such a configuration cannot preserve $A$ and $B$. On the other hand, when $A<0$ and $B<0$, the modes with $v\approx-1$ cannot flip sign. We will see this pattern in our numerical calculations.

\begin{figure}[!t]
\begin{centering}
	\includegraphics[width=0.95\columnwidth]{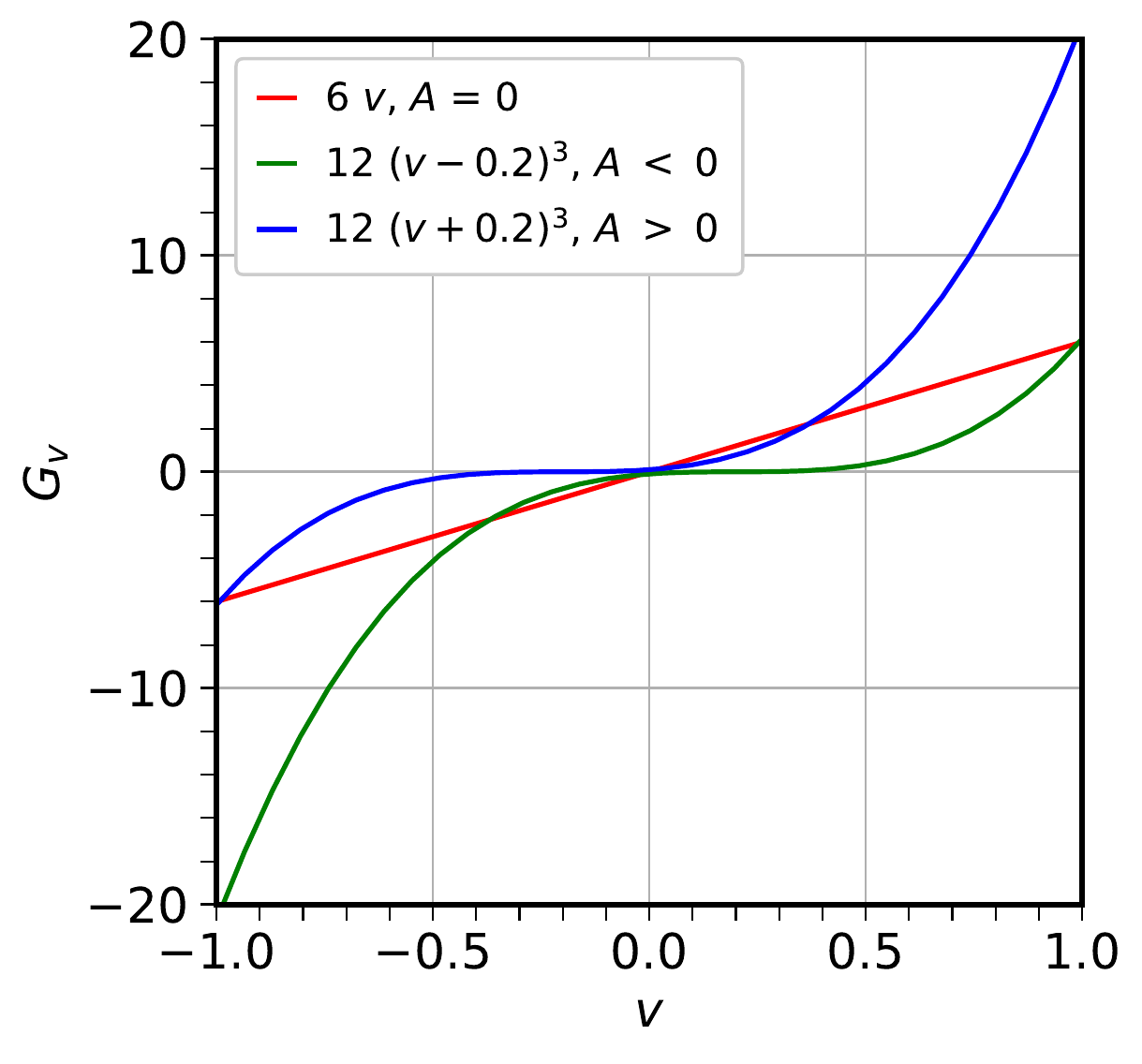}
\end{centering}	
\caption{ELN distributions used in our numerical examples. The case $A=0$ has vanishing lepton asymmetry, whereas $A<0$ and $A>0$ have more net negative and positive lepton number, respectively.}
\label{fig1}
\end{figure}
\begin{figure*}[t!]
\hspace{0.1cm}	\includegraphics[width=0.64\columnwidth]{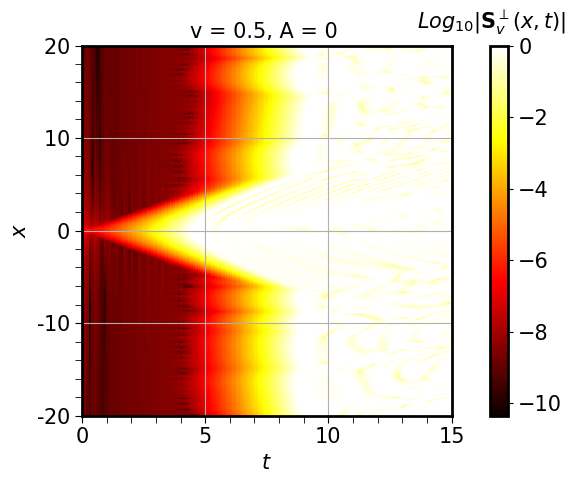}
\hspace{0.1cm}	\includegraphics[width=0.64\columnwidth]{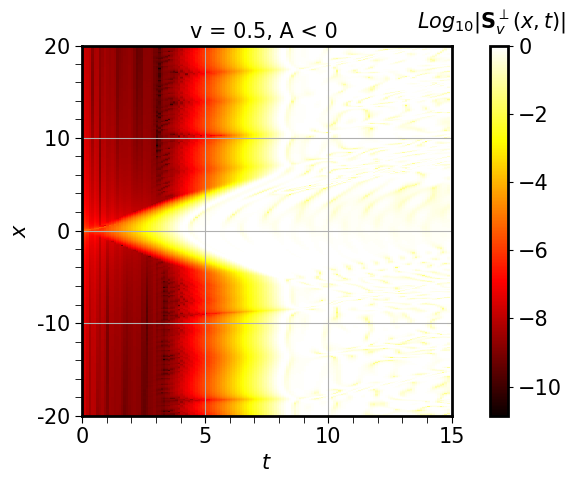}
\hspace{0.1cm}	\includegraphics[width=0.64\columnwidth]{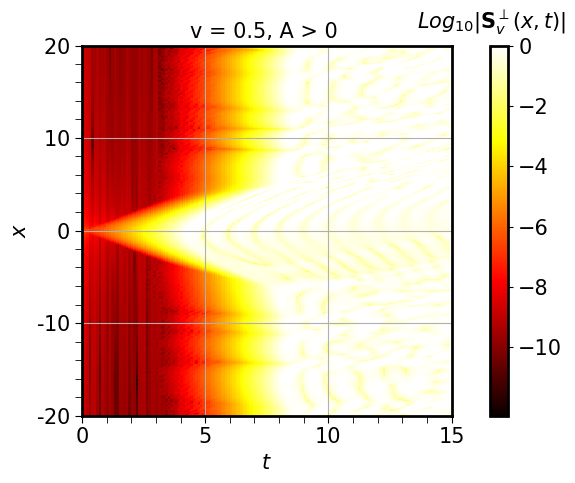}\\
	\includegraphics[width=0.66\columnwidth]{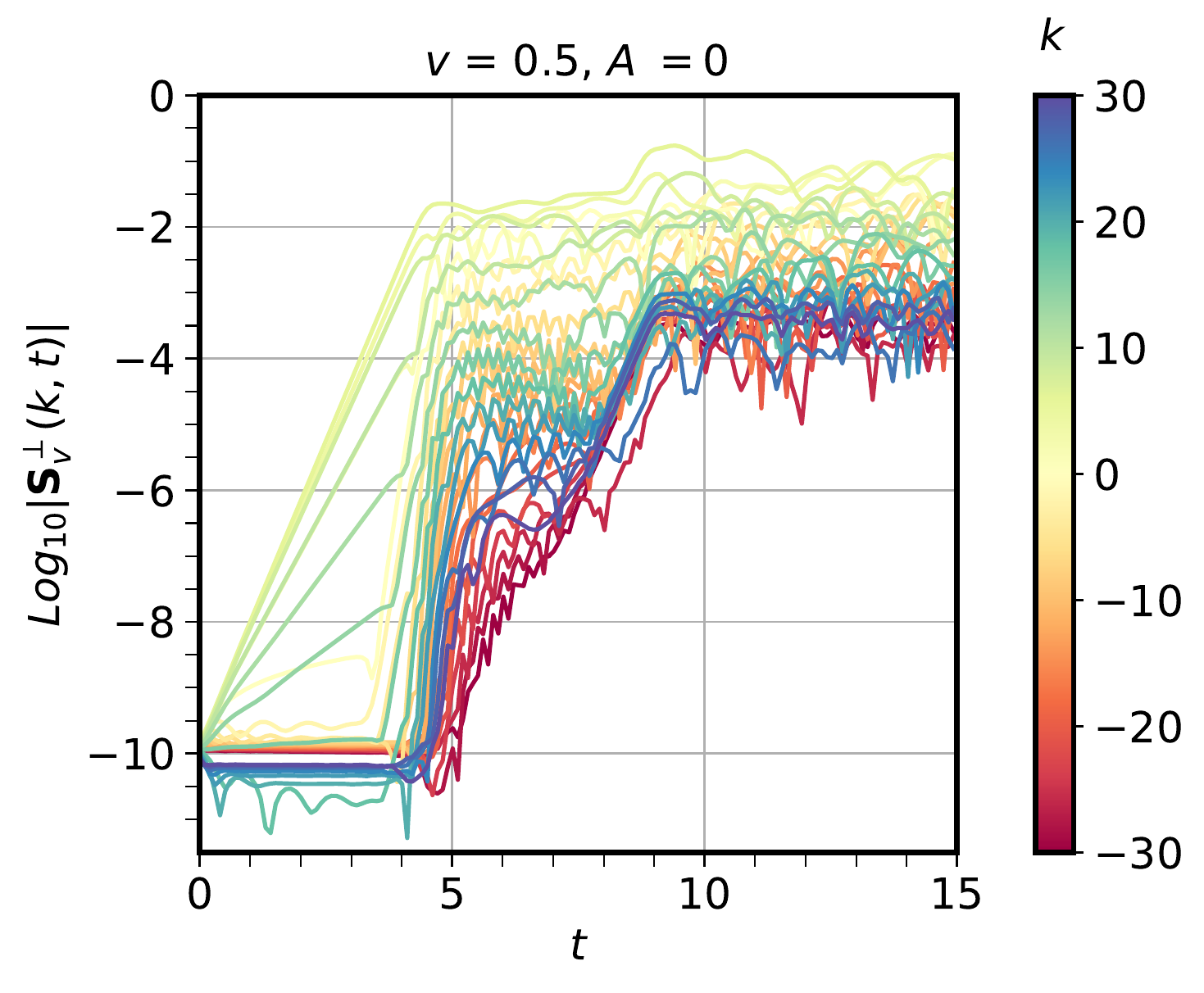}
	\includegraphics[width=0.66\columnwidth]{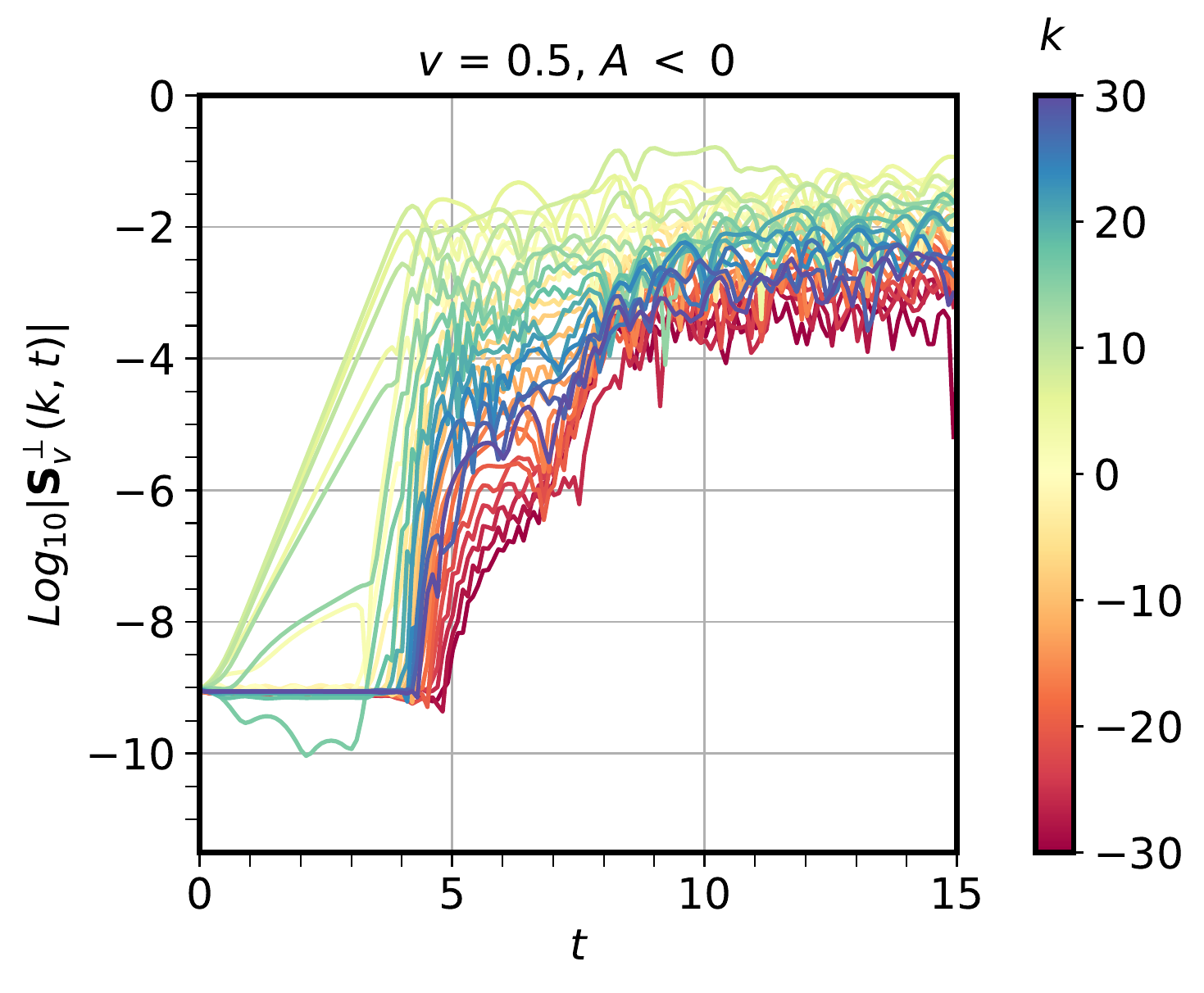}
	\includegraphics[width=0.66\columnwidth]{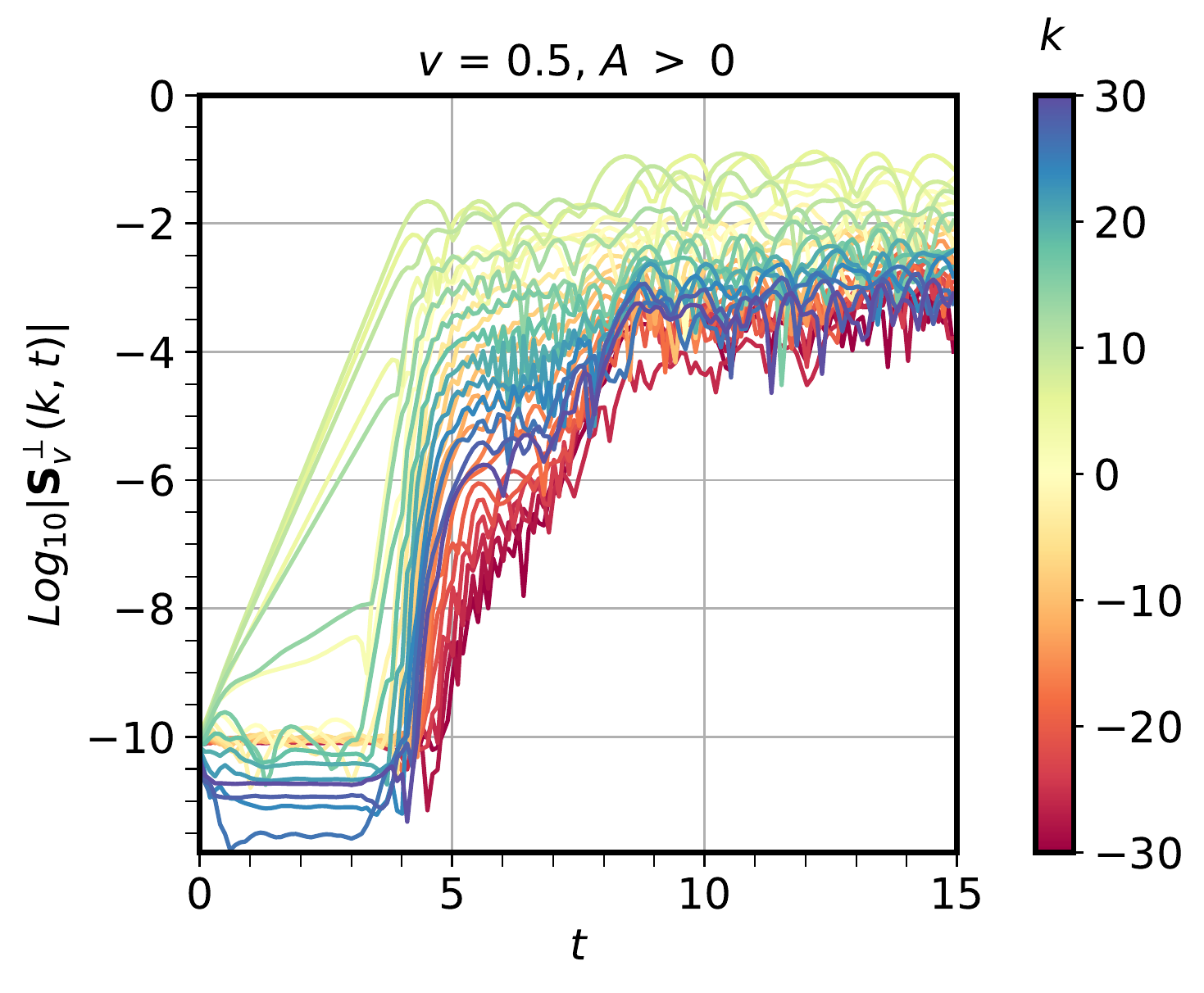}
	\caption{Top: Absolute value of the off-diagonal element of the density matrix, color-coded as per the color-bar, shown for the velocity mode $v=0.5$. Bottom: Growth of the off-diagonal elements of the density matrix as a function of time for various different $k$ modes, color-coded as per the color-bar, for the same velocity mode. Left panels show the case $A=0$, middle panels $A<0$, and right panel $A>0$. Note the approximately steady state solution at late times.}
	\label{fig2}
\end{figure*}

\begin{figure*}[!h]
	\includegraphics[width=0.66\columnwidth]{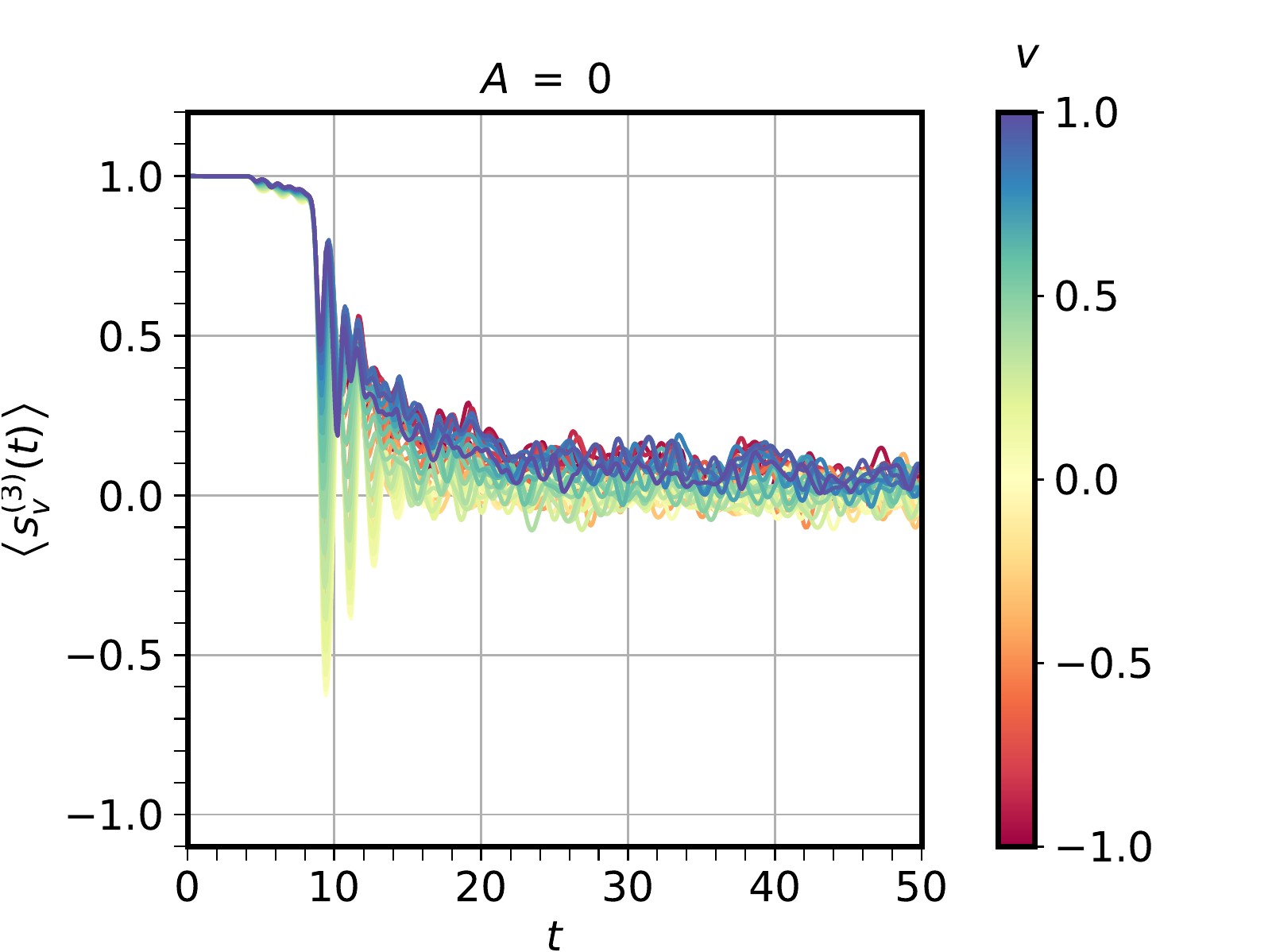}
	\includegraphics[width=0.66\columnwidth]{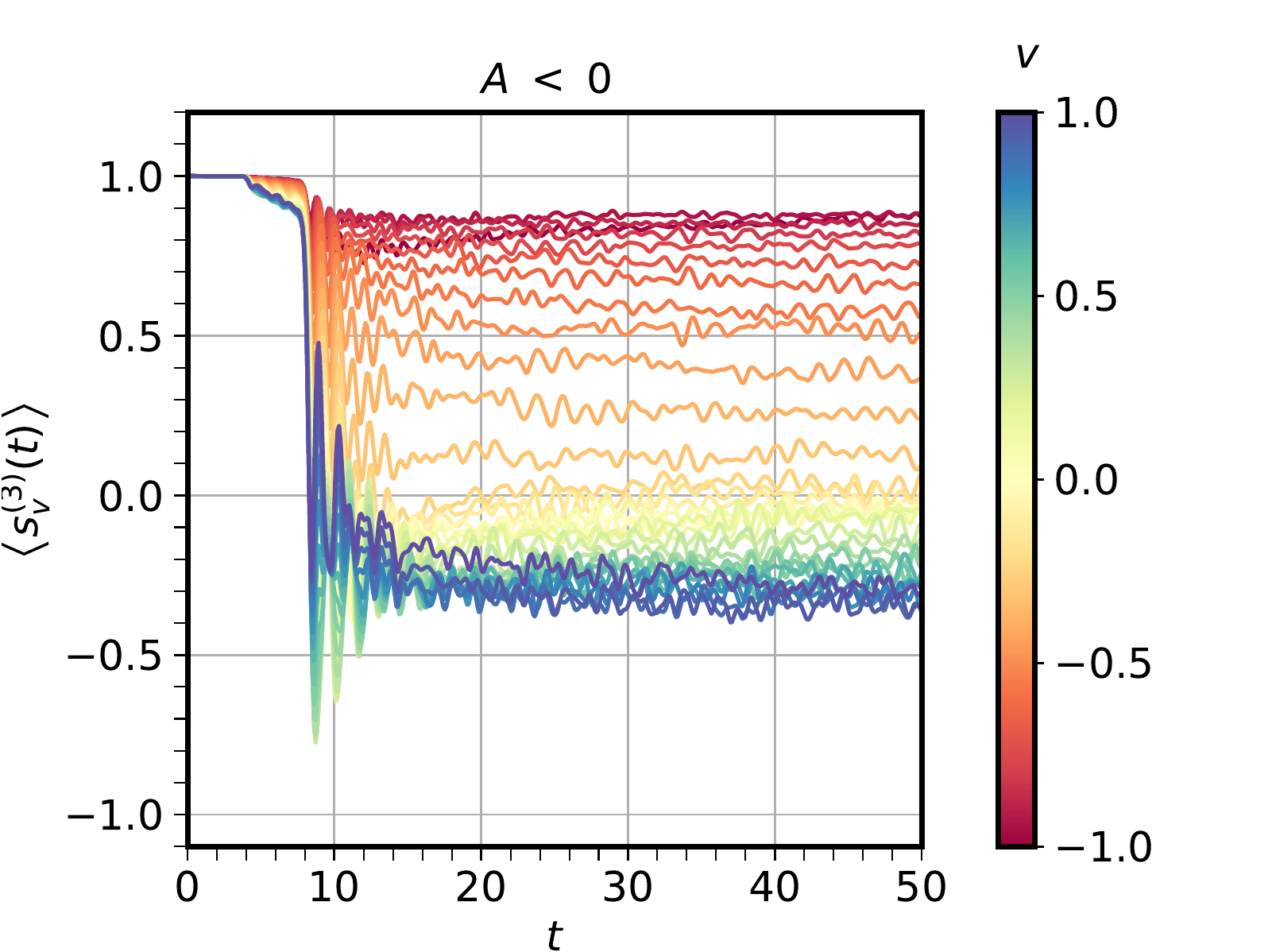}
	\includegraphics[width=0.66\columnwidth]{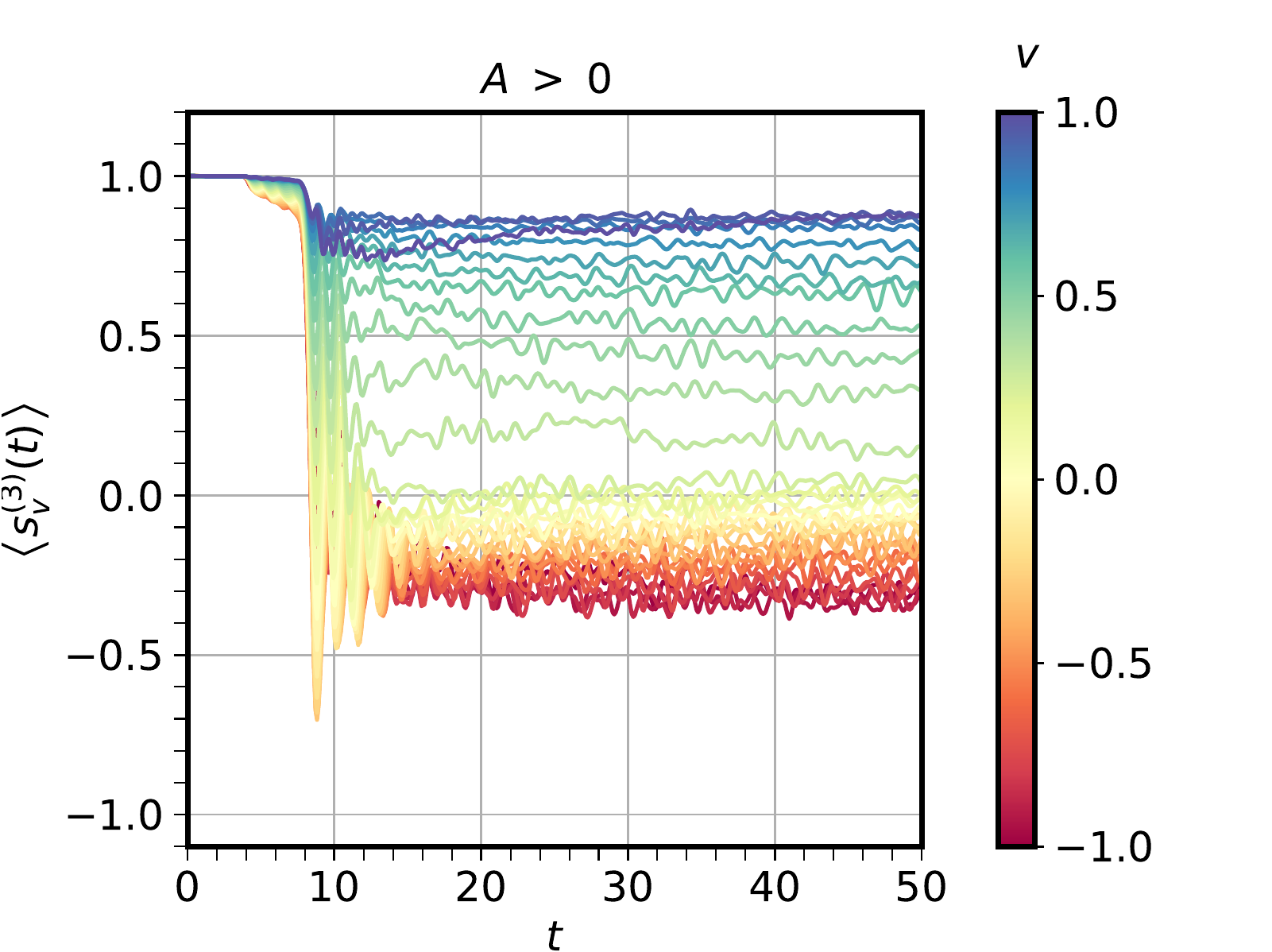}\\
	\includegraphics[width=0.66\columnwidth]{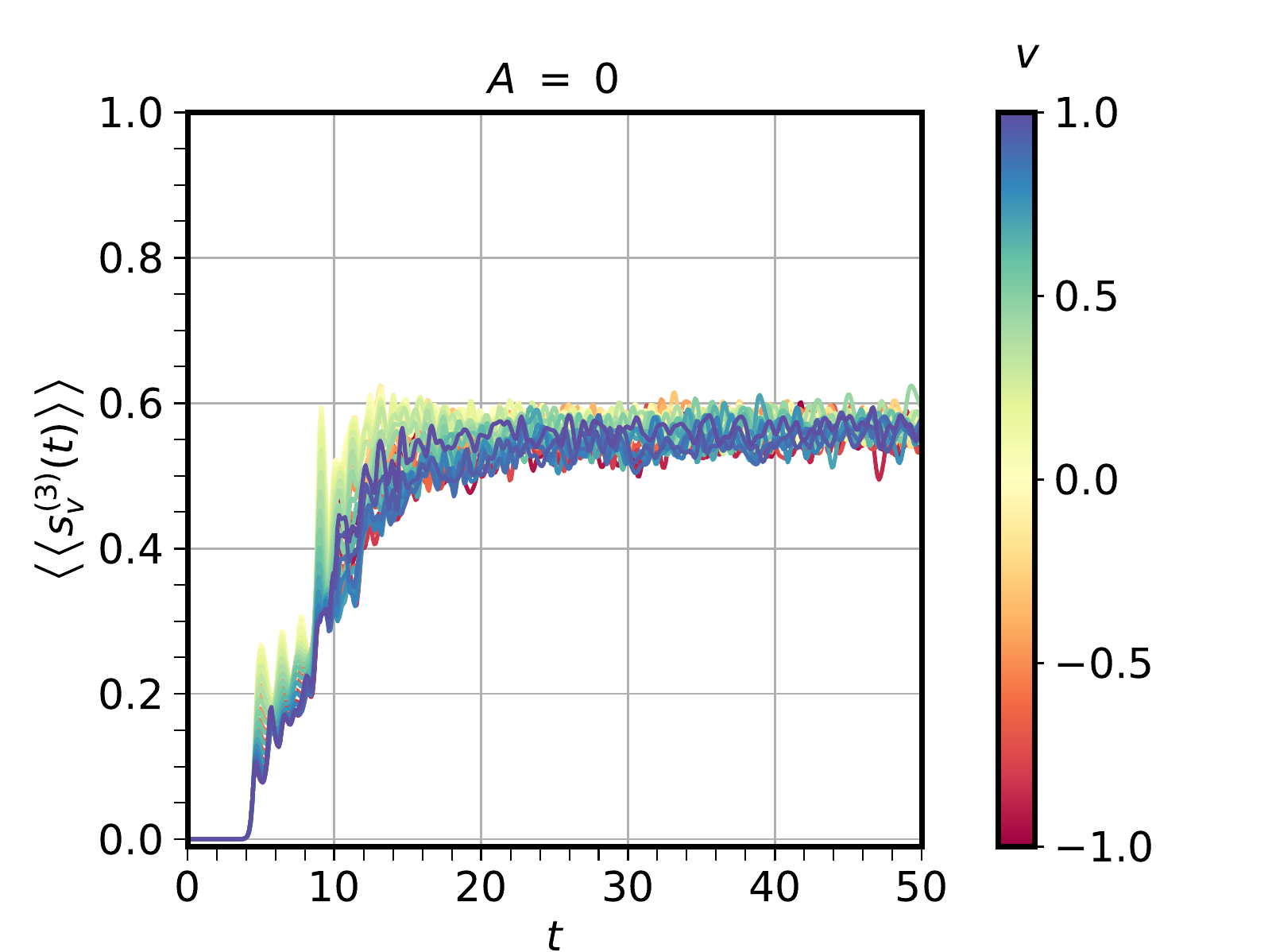}
	\includegraphics[width=0.66\columnwidth]{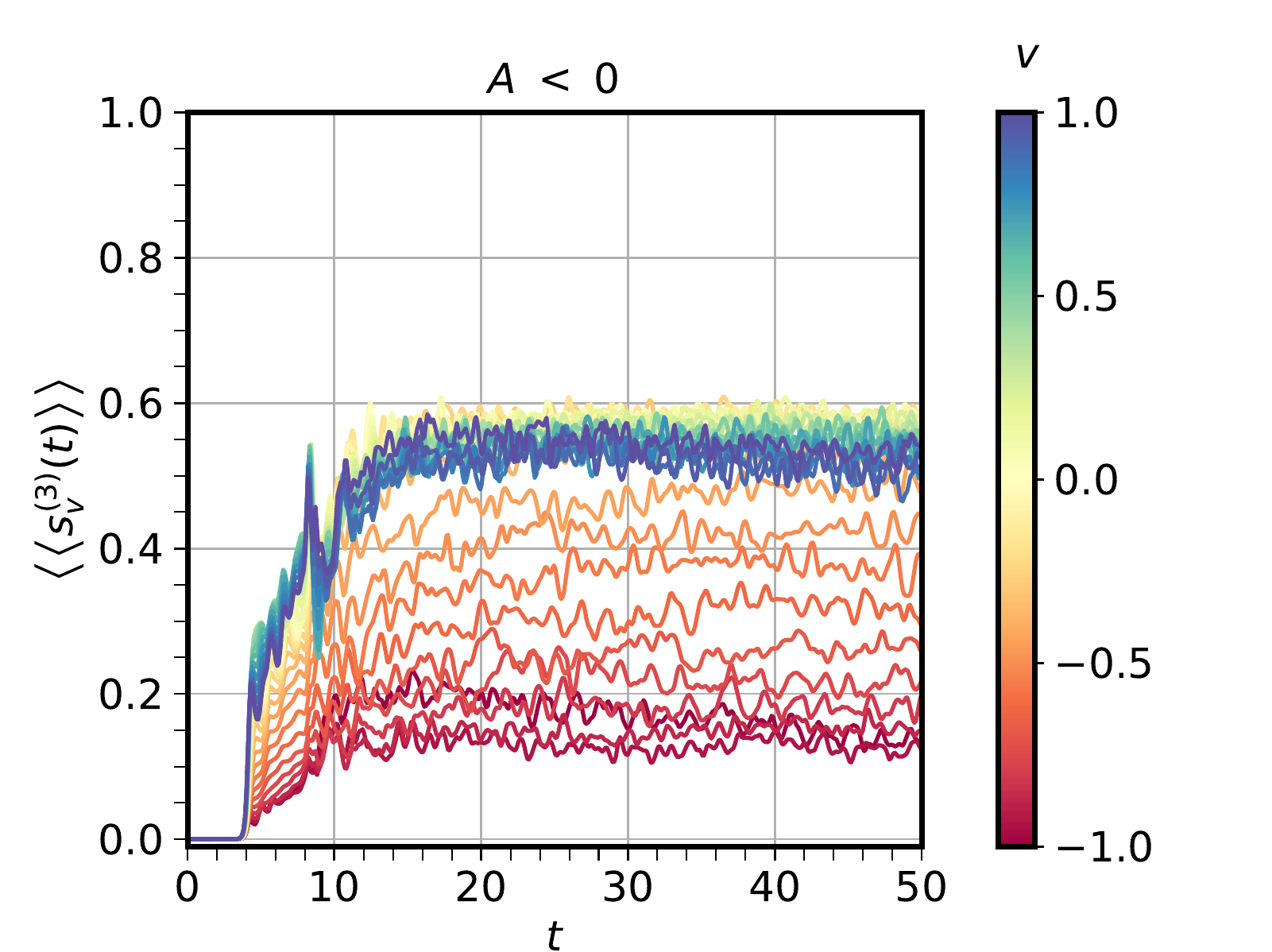}
	\includegraphics[width=0.66\columnwidth]{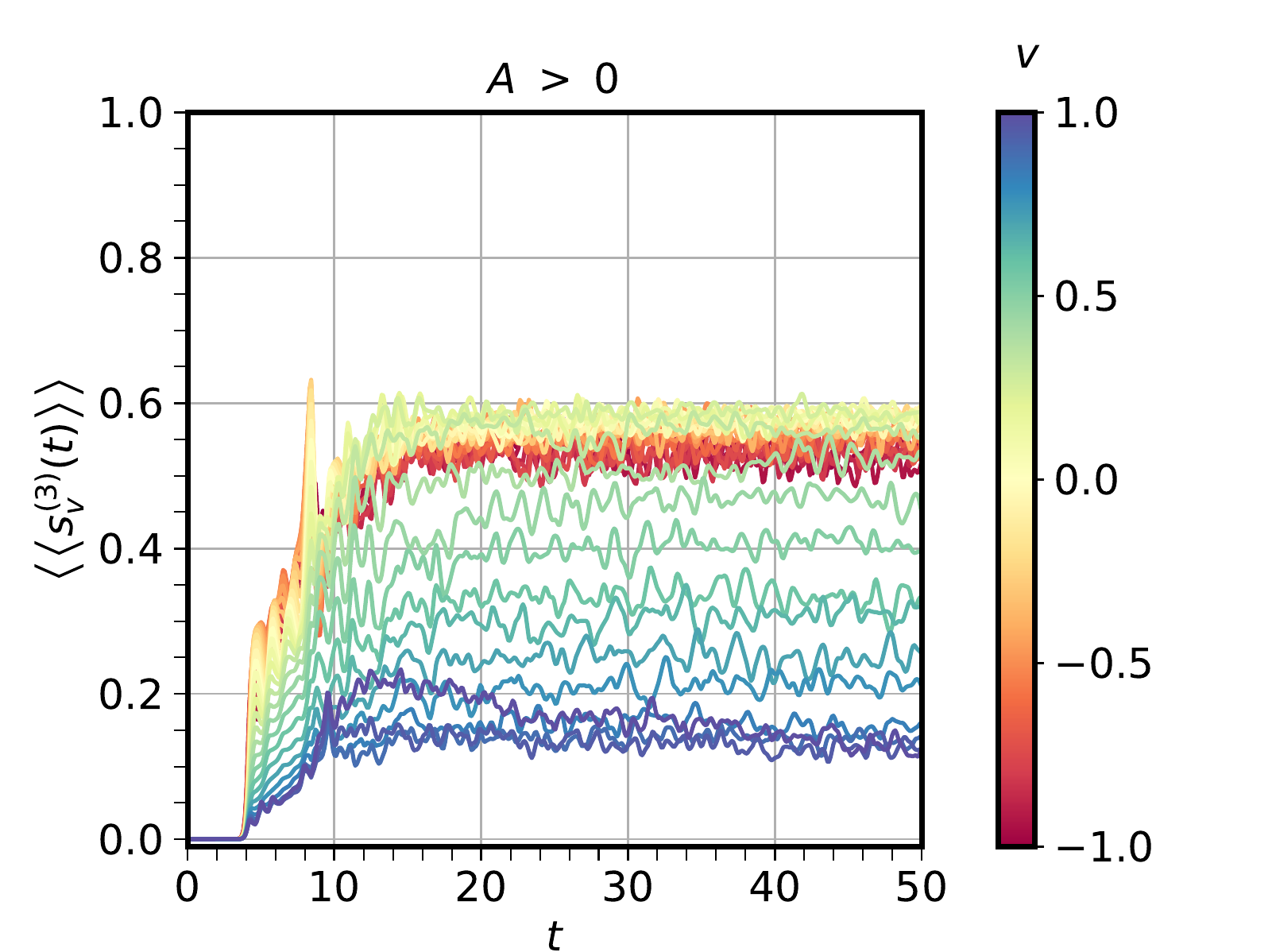}\\
	\caption{Top: Mean value of spatial distributions of $s_{v}^{(3)}$ varying with time, for a set of velocity modes color-coded by the velocity. Bottom: Standard deviation of the spatial distributions of $s_{v}^{(3)}$ as a function of time. Left panels show the case $A=0$, middle panels $A<0$, and right panel $A>0$. Note that both the spatial mean and standard deviation become approximately stationary in time in the nonlinear regime. Note also the spectral-swap-like separation of $s_{v}^{(3)}$ into two cohorts for $A\neq0$.} 
	\label{fig3}
\end{figure*}

\begin{figure*}[!t]
	\hspace{-0.2cm}\includegraphics[width=0.69\columnwidth]{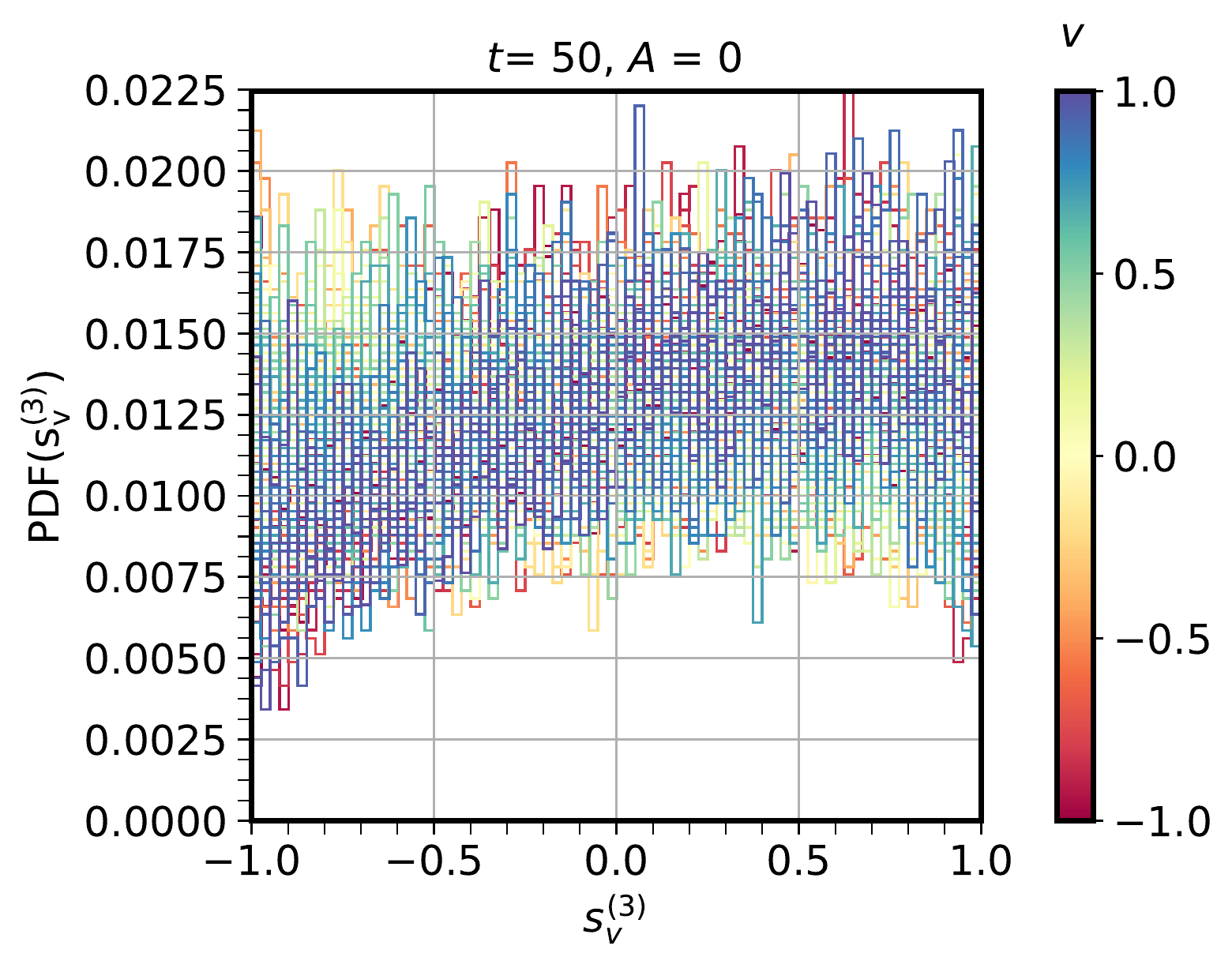}
	\includegraphics[width=0.66\columnwidth]{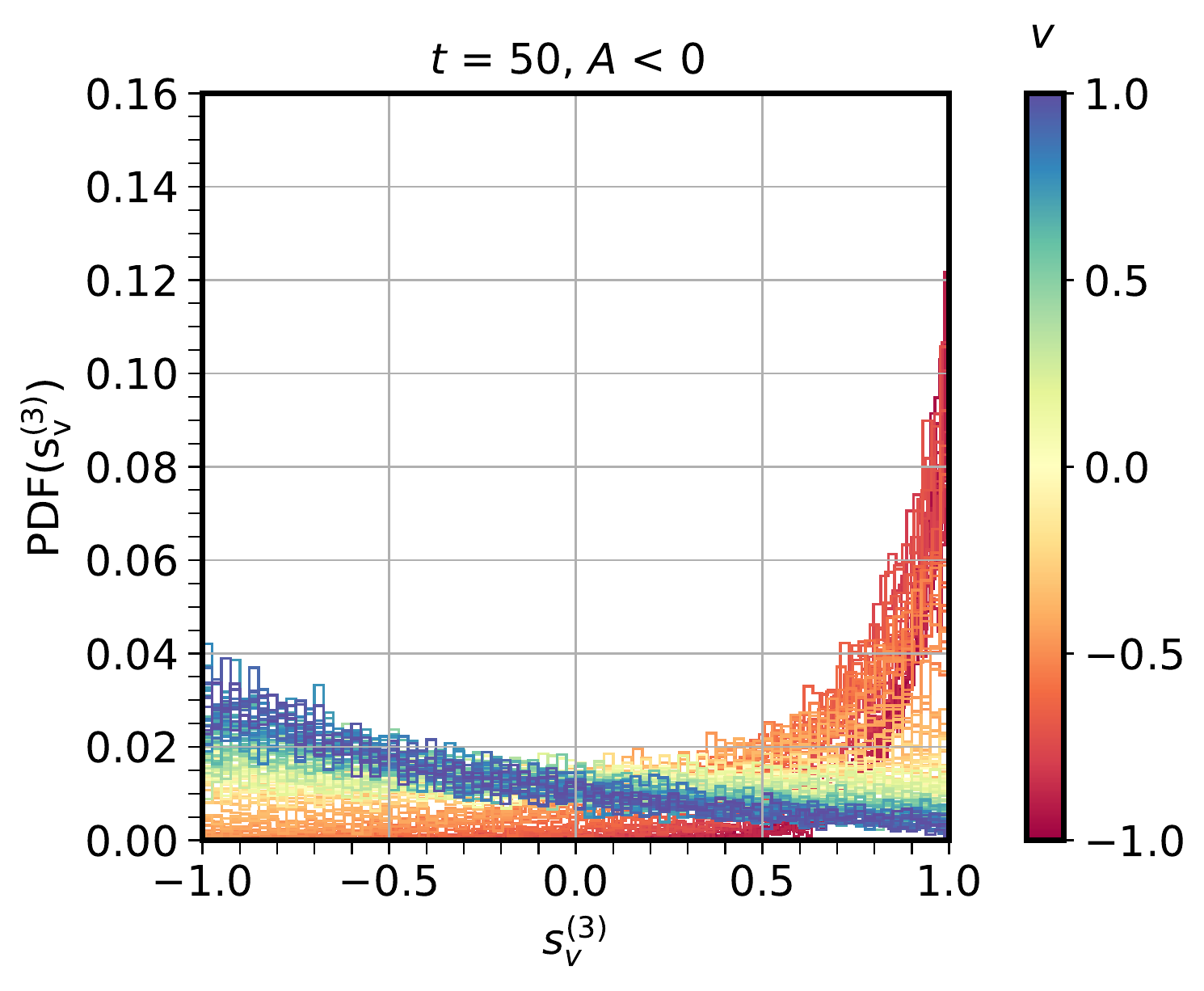}
	\includegraphics[width=0.66\columnwidth]{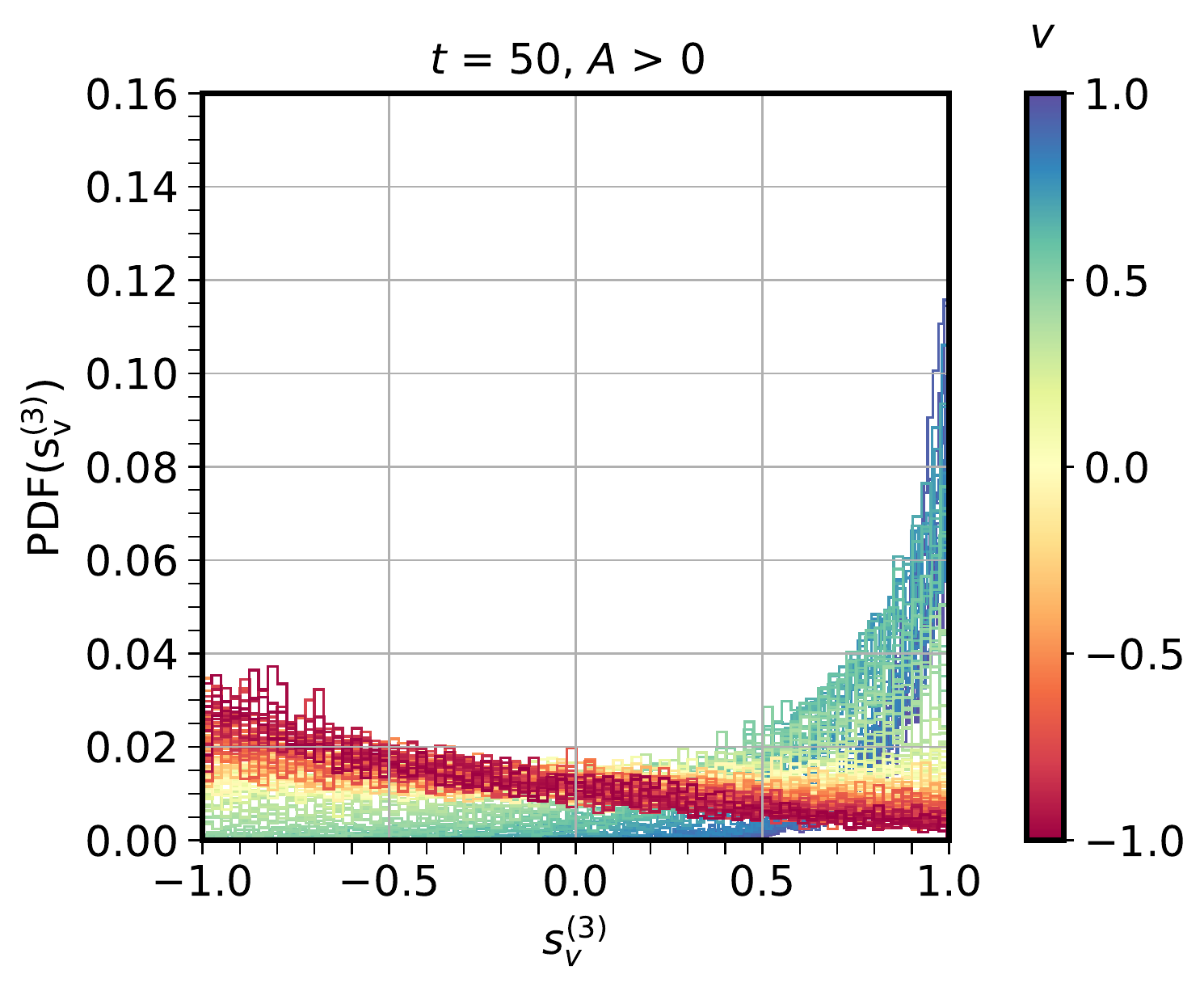}\\
	\includegraphics[width=0.66\columnwidth]{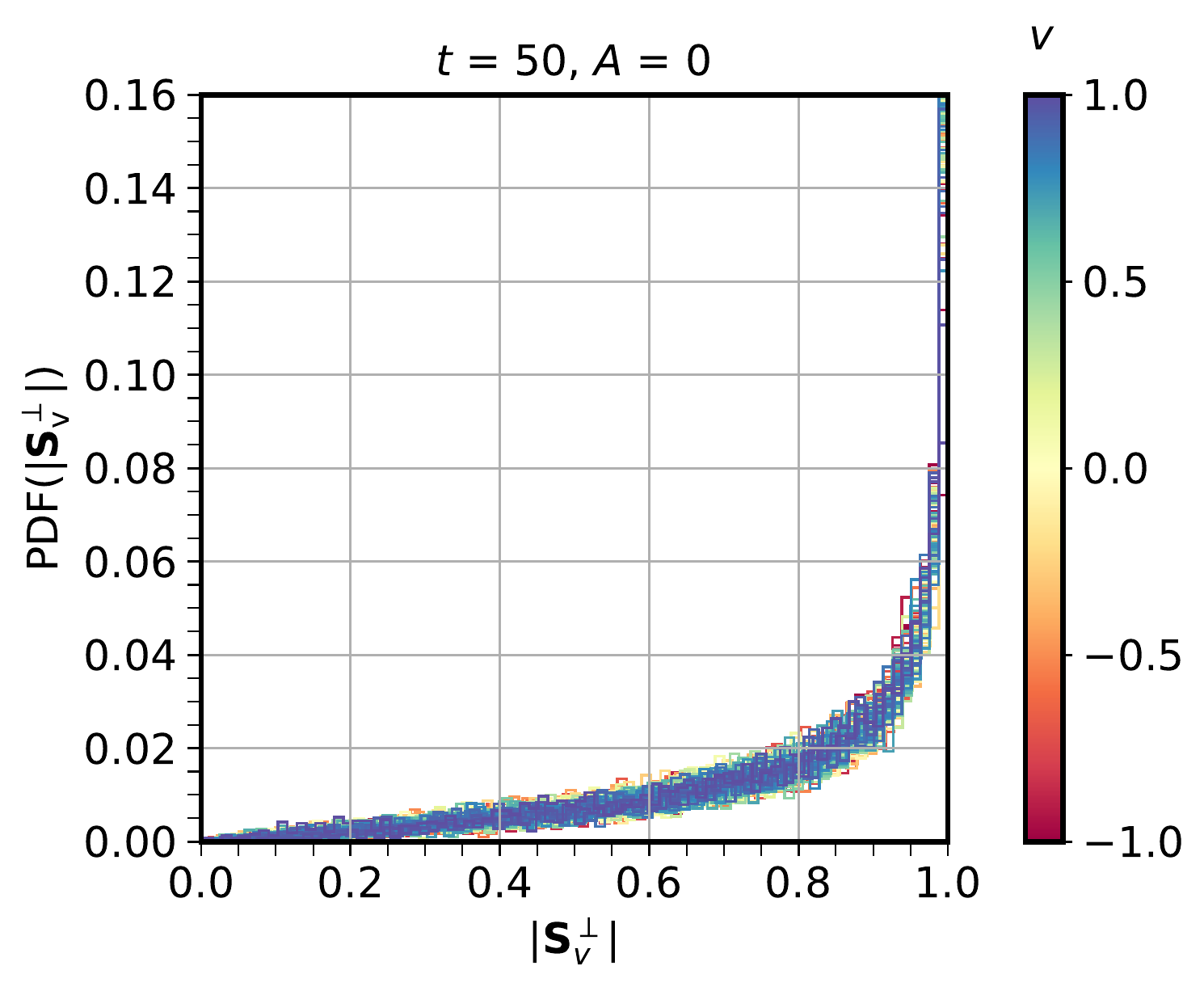}
	\includegraphics[width=0.66\columnwidth]{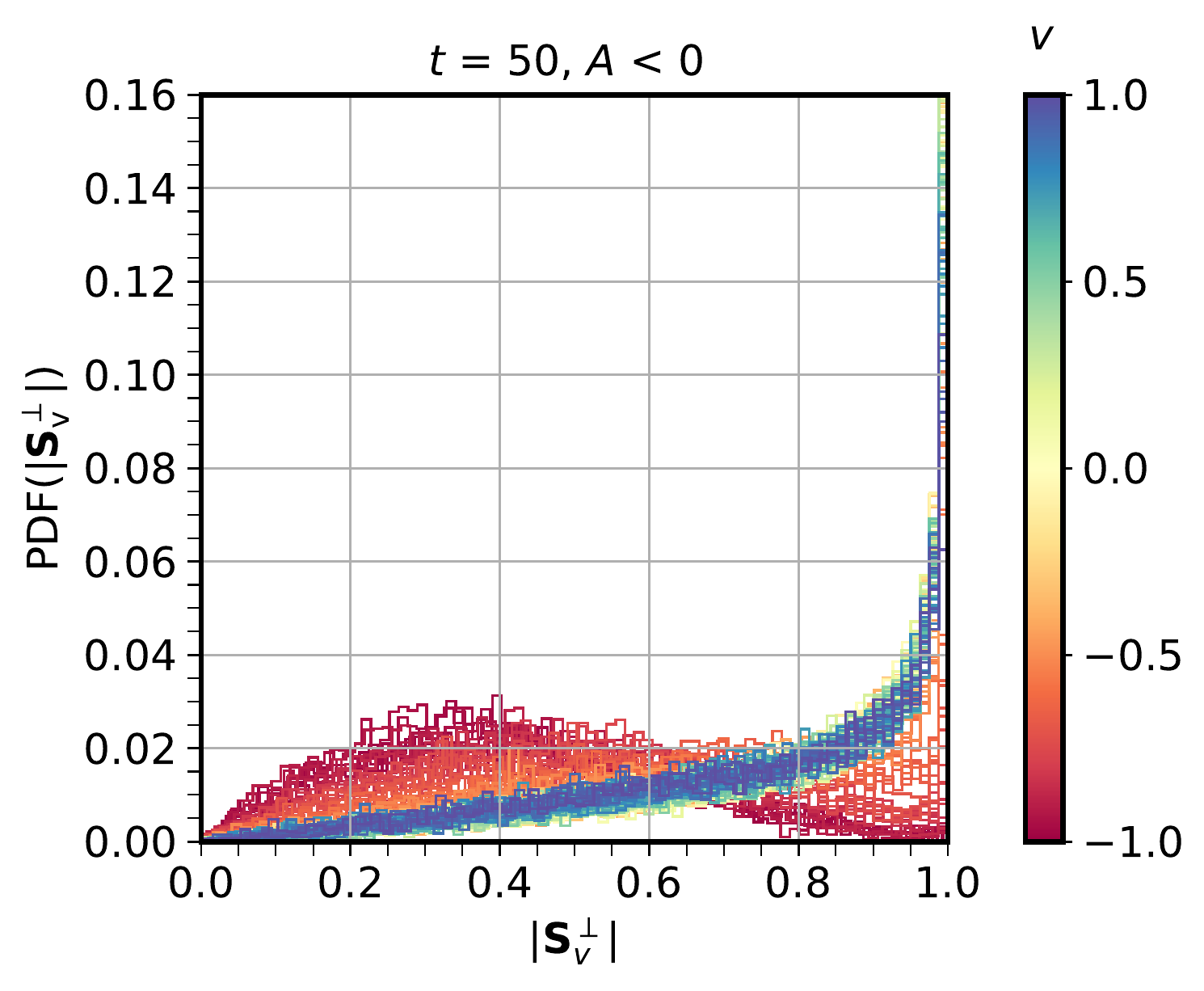}
	\includegraphics[width=0.66\columnwidth]{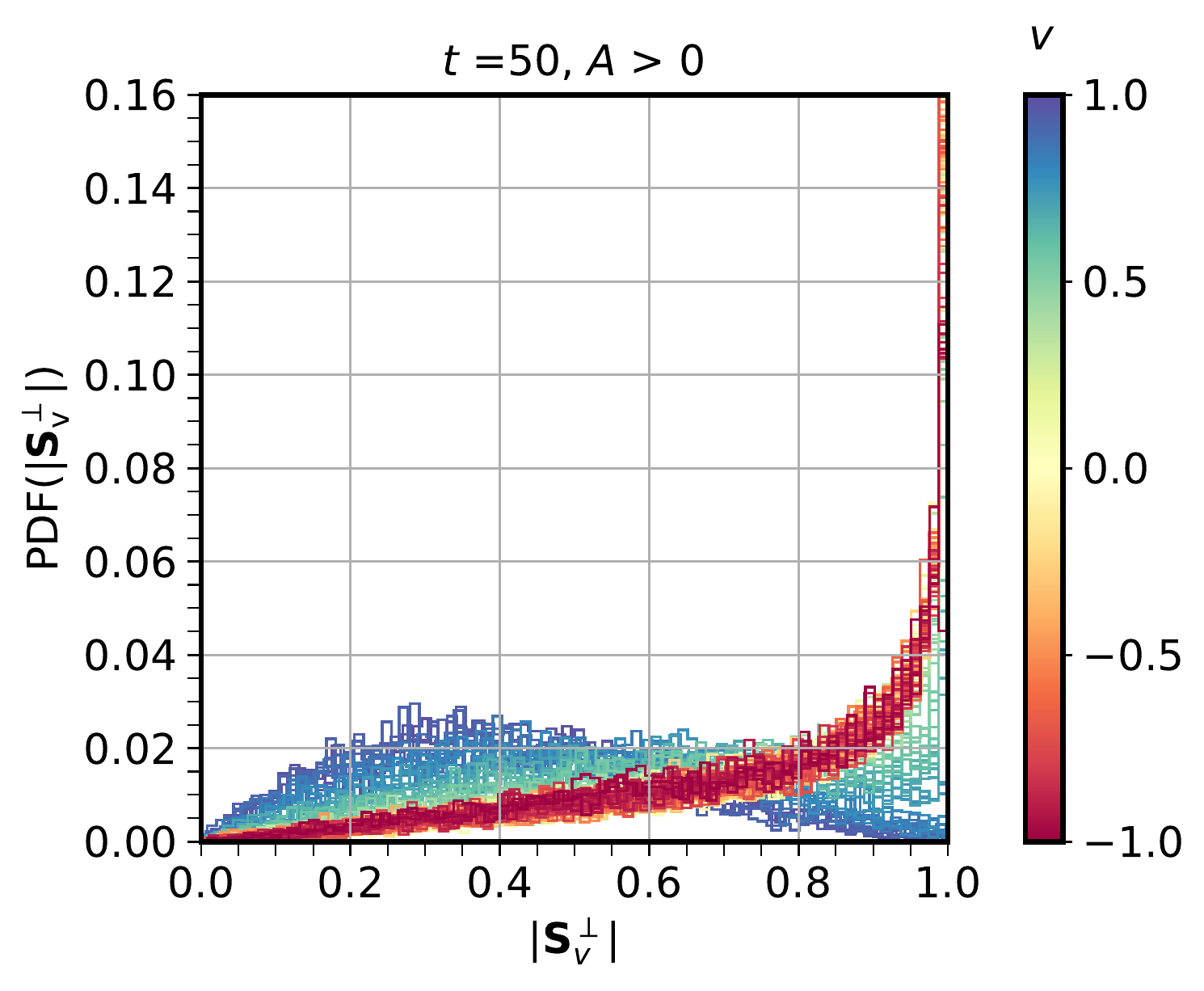}
	\caption{Top: Histograms of spatial distributions of $s_{v}^{(3)}$ at time $t = 50$ for different velocity modes color-coded by the velocity. Bottom: Same for histograms for $|\mathbf{S}_{v}^{{\perp}}|$.} 
	\label{fig4}
\end{figure*}
\begin{figure*}[!t]
	\includegraphics[width=0.6\columnwidth]{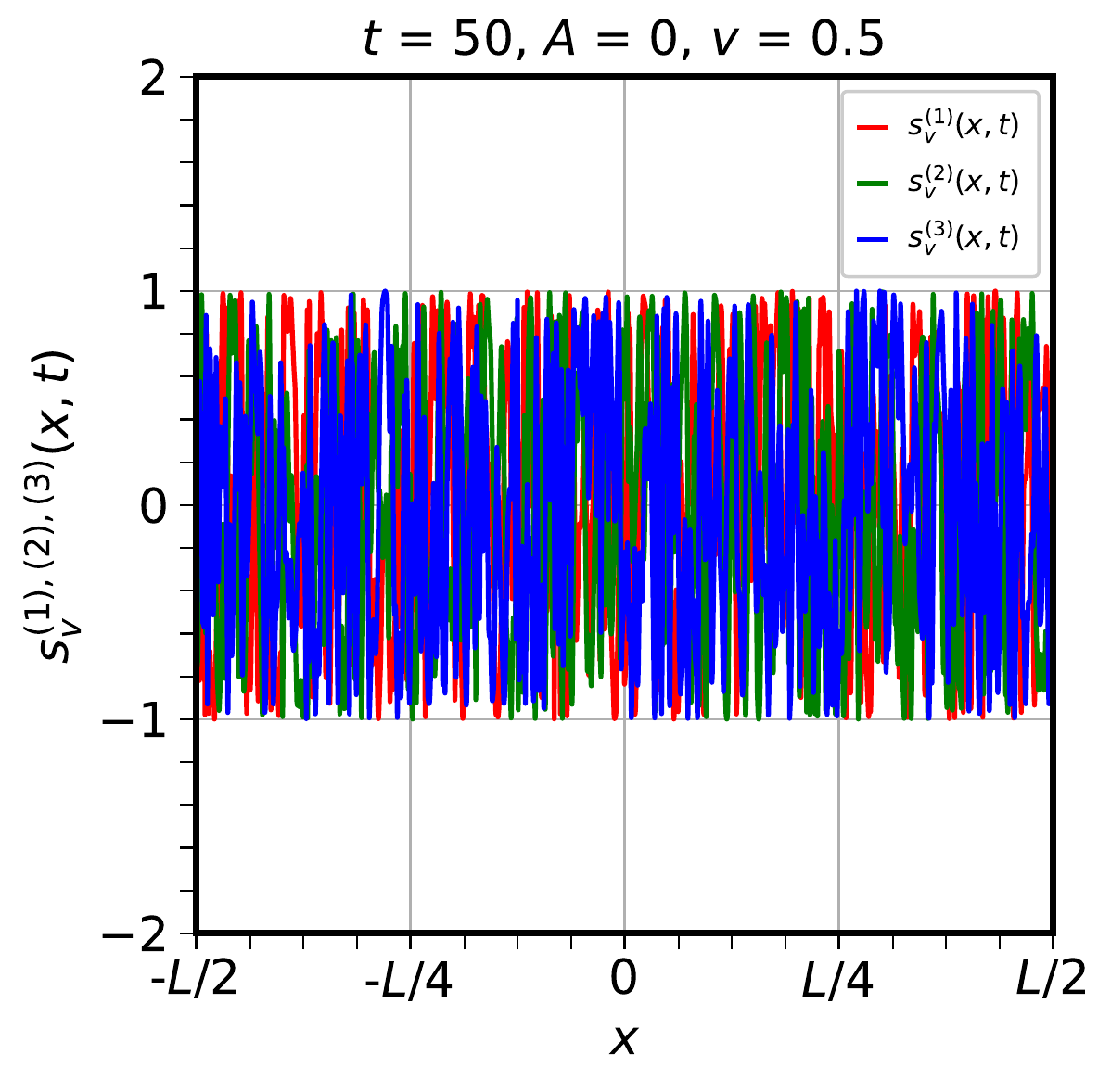}
	\includegraphics[width=0.6\columnwidth]{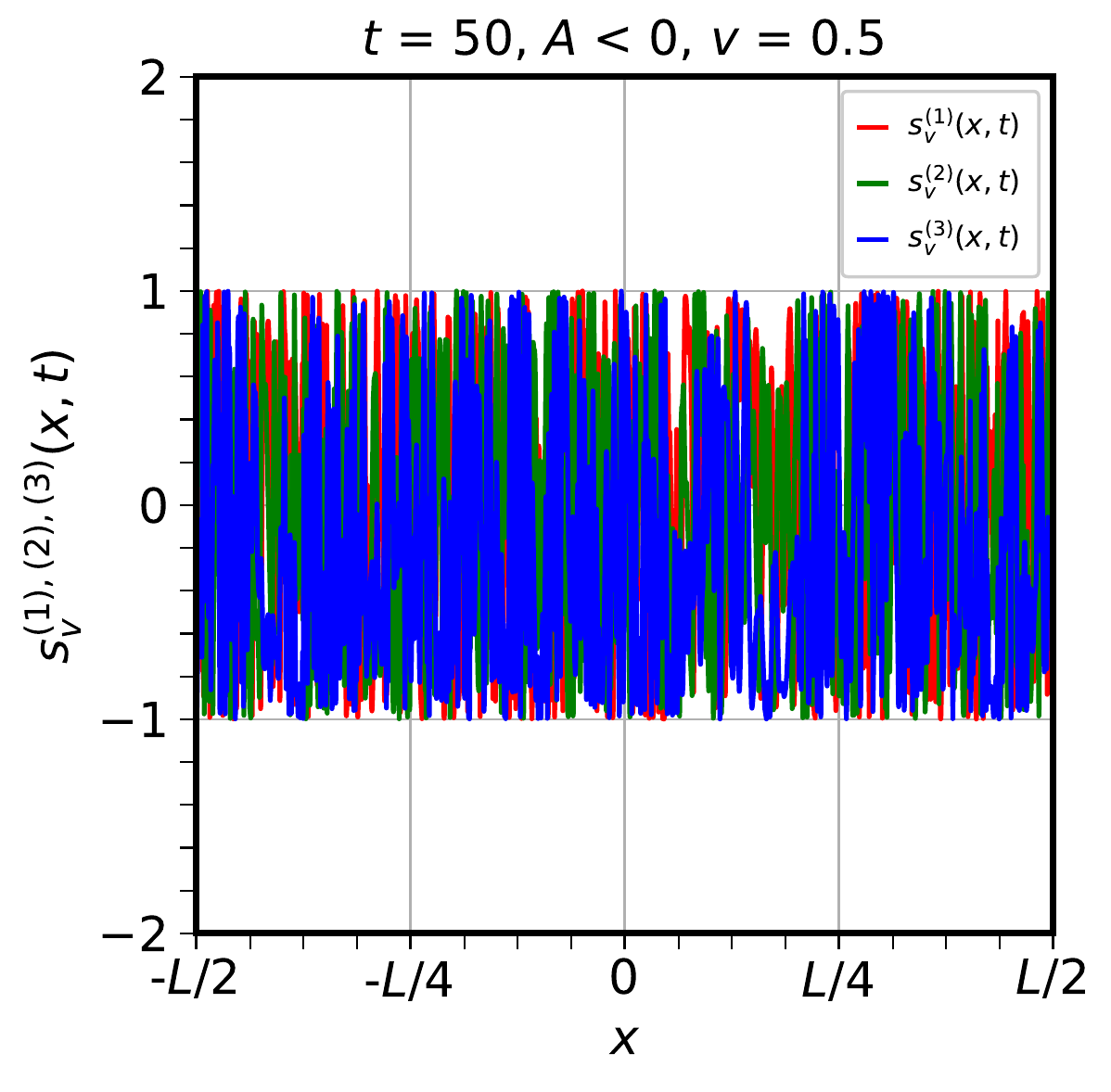}
	\includegraphics[width=0.6\columnwidth]{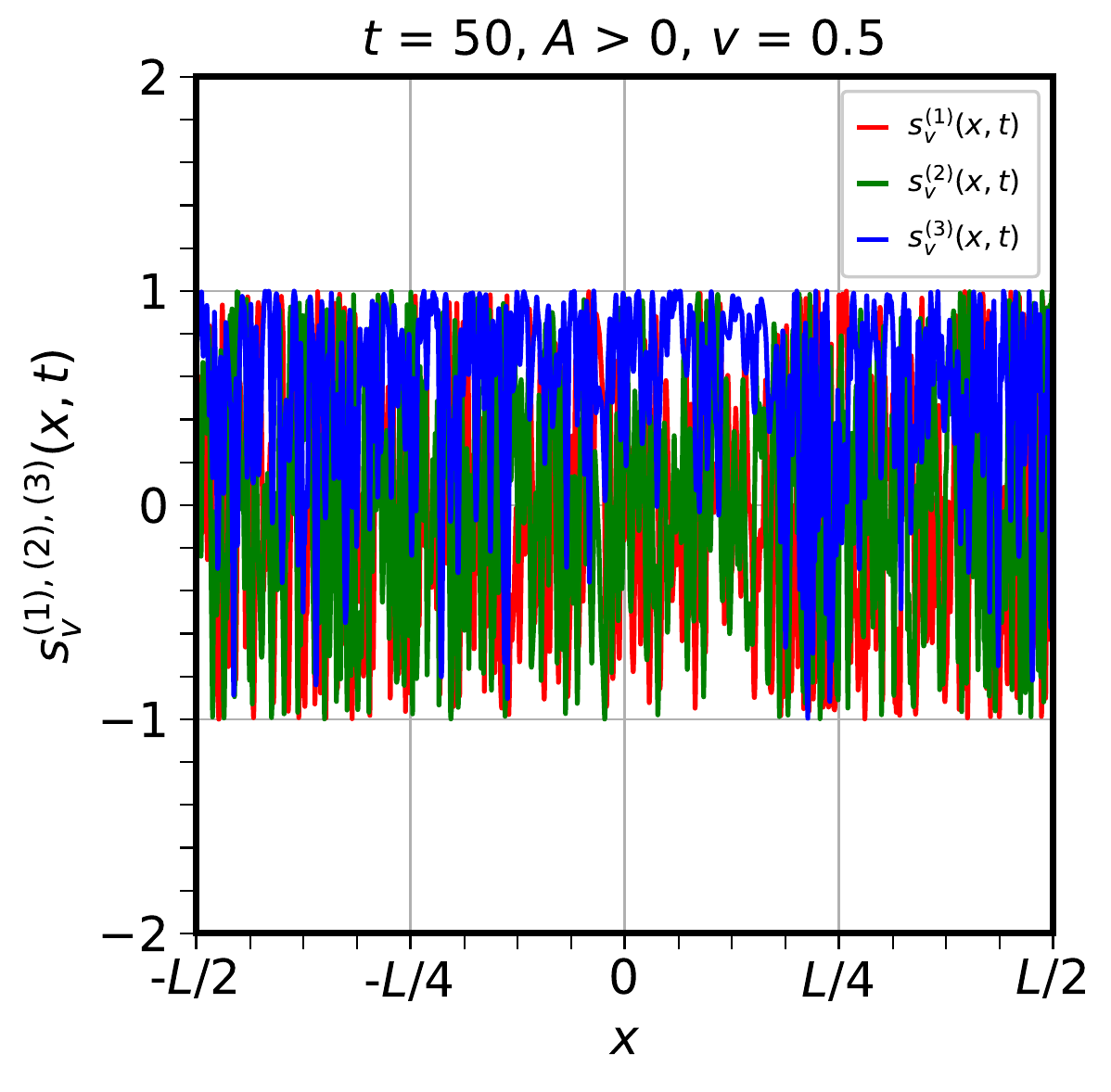}\\
	\includegraphics[width=0.6\columnwidth]{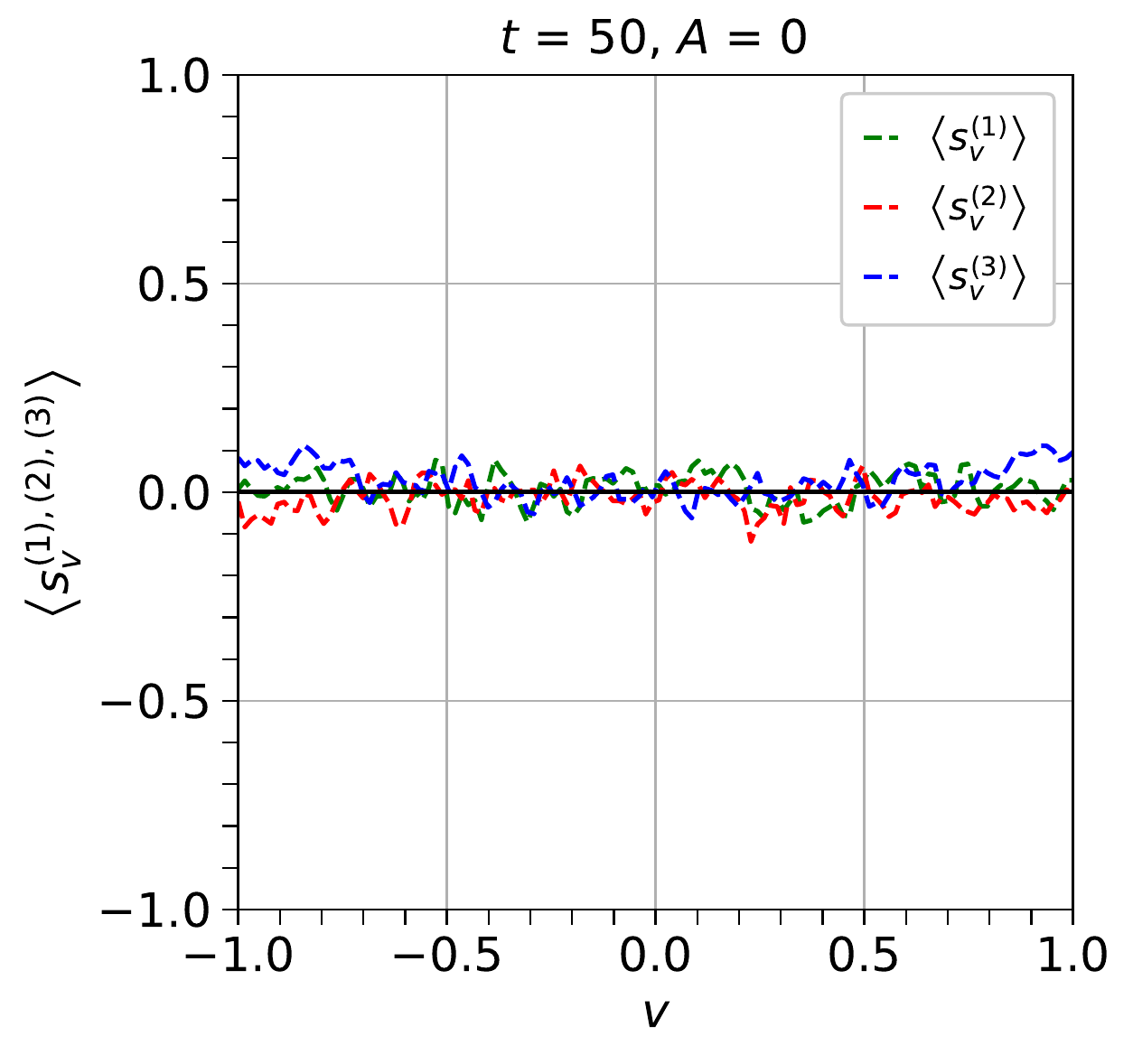}
	\includegraphics[width=0.6\columnwidth]{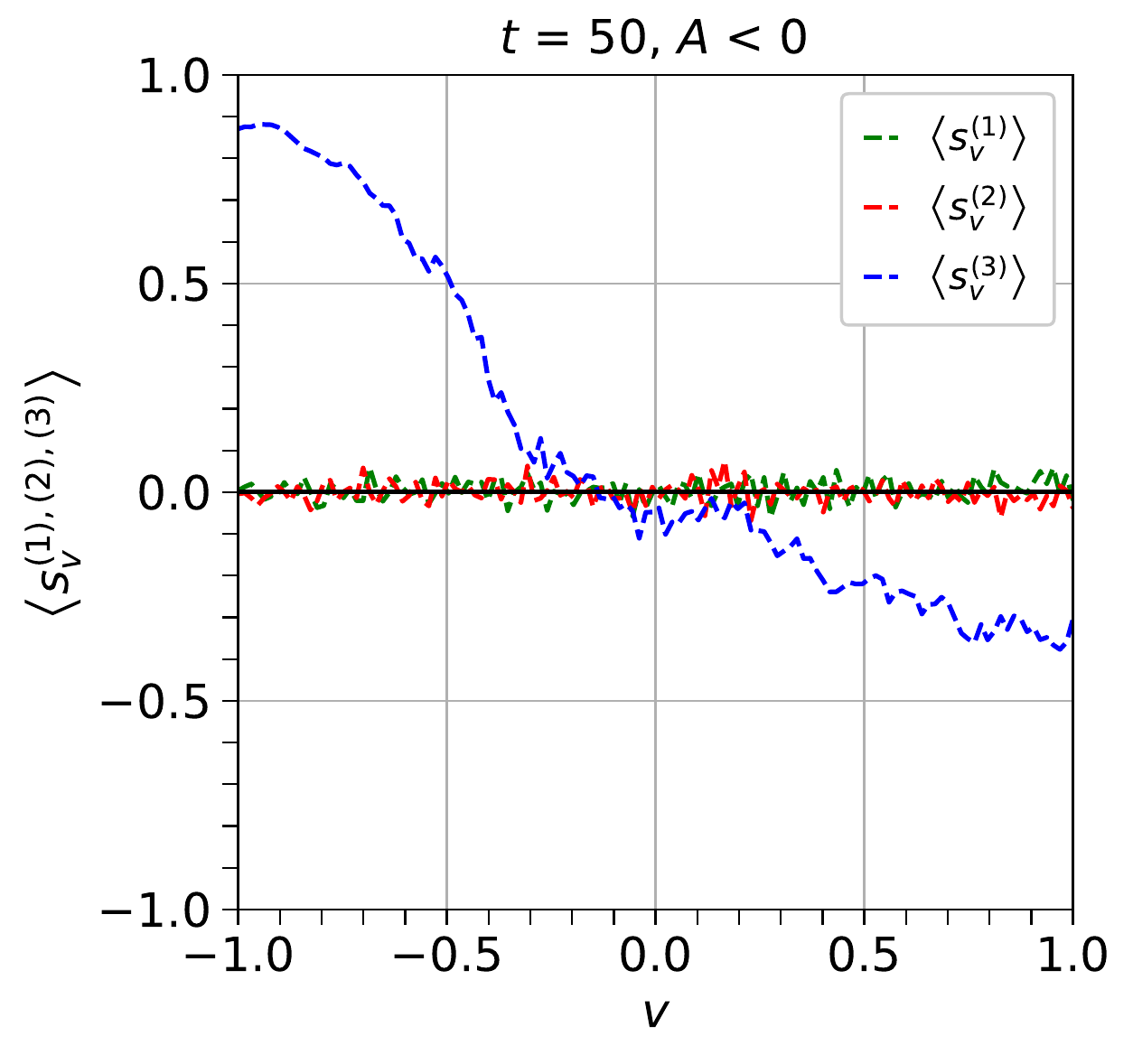}
	\includegraphics[width=0.6\columnwidth]{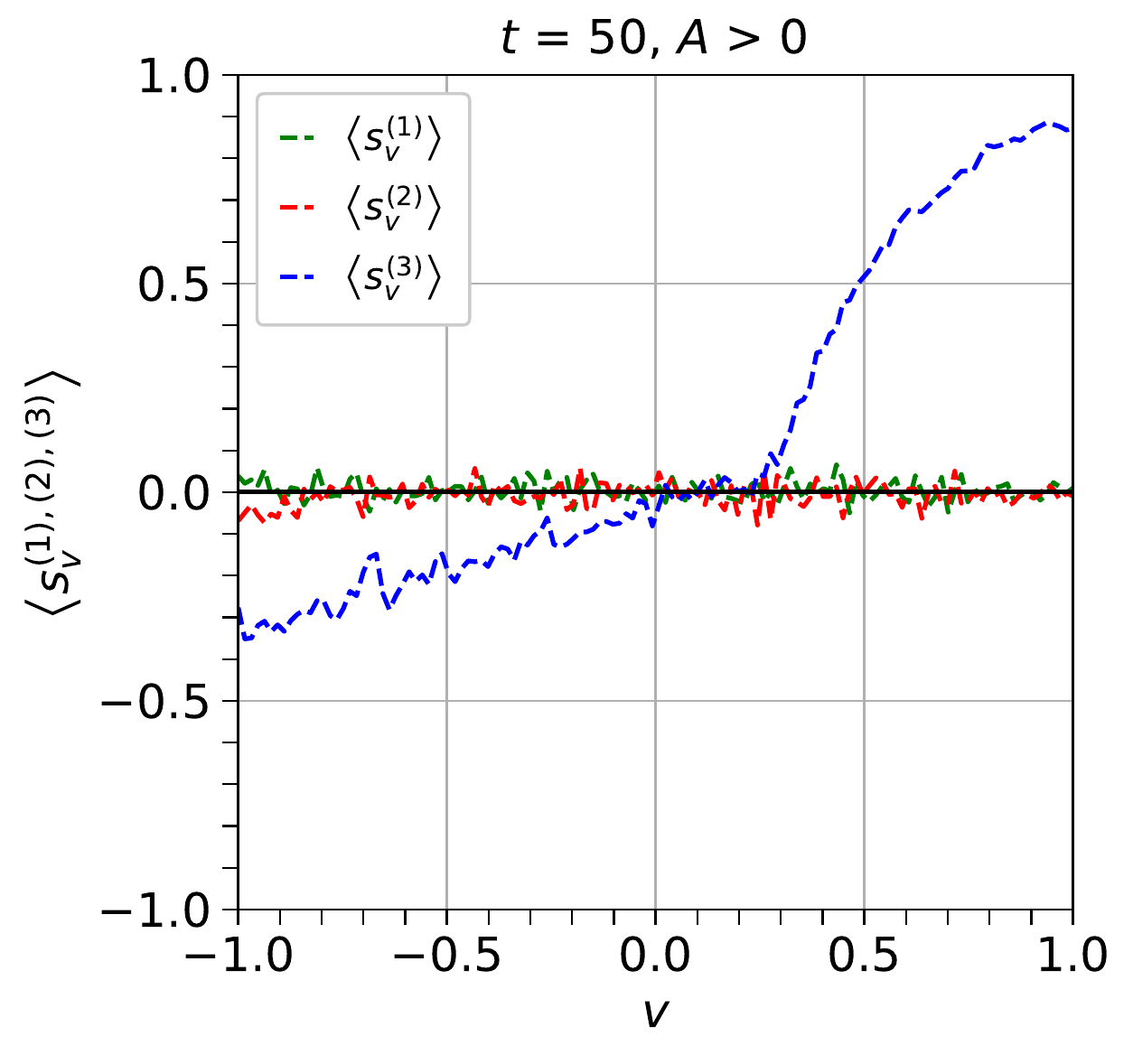}
	\caption{Top: Spatial variation of the three components of the polarization vector at time $t = 50$ for $v=0.5$. Bottom: Spatially averaged value of the three components of the polarization vector as a function of different velocity modes at time $t = 50$. Left panels show the case $A=0$, middle panels $A<0$, and right panel $A>0$. Note the fast oscillations over spatial coordinate, but a relatively simpler velocity dependence for the spatially averaged solution. For $A=0$ there is flavor depolarization, whereas there is a partial depolarization, with an incomplete swap, for $A\neq0$.}
	\label{fig5}
\end{figure*}

\begin{figure*}[!t]
	\hspace{0.3cm}\includegraphics[width=0.62\columnwidth]{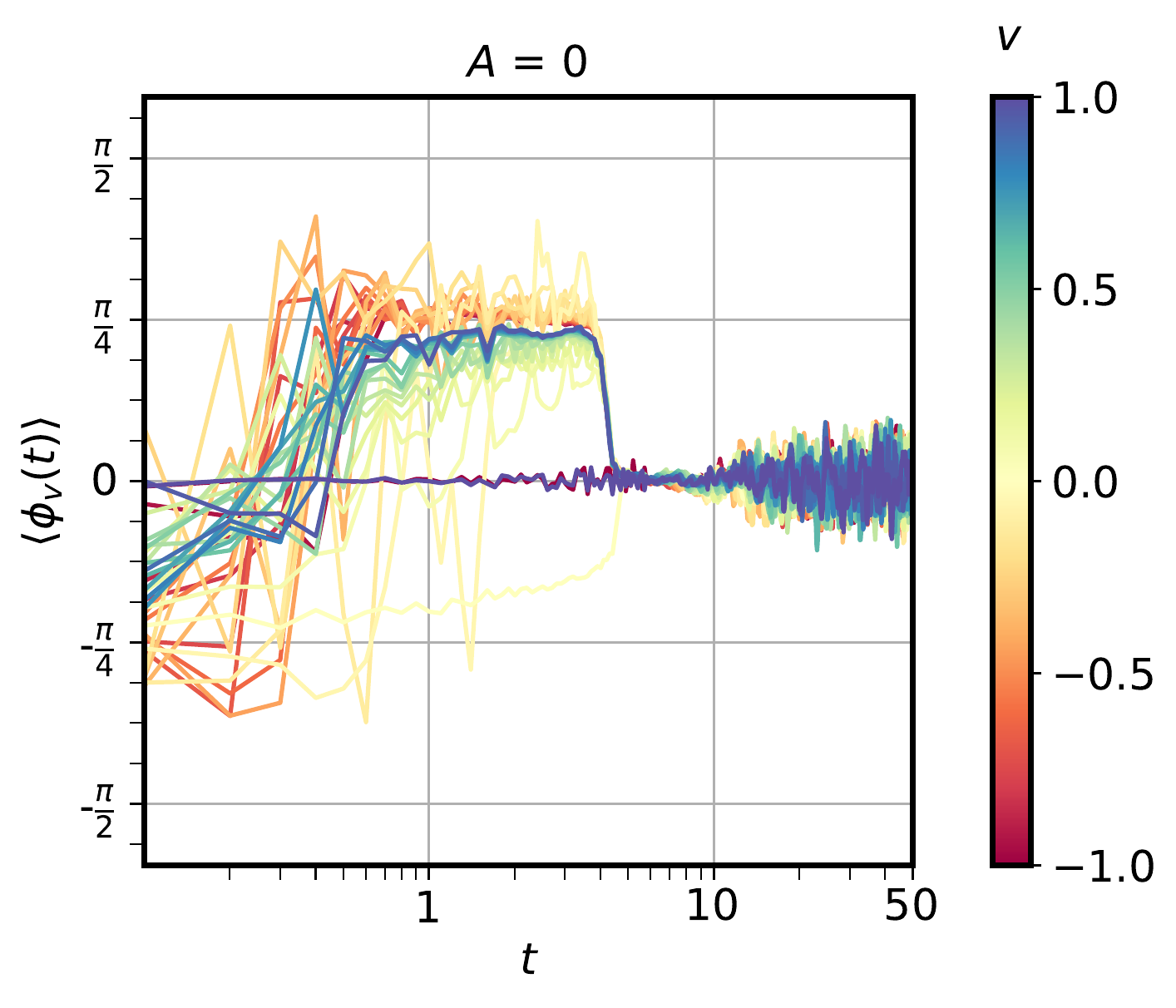}
	\hspace{0.3cm}\includegraphics[width=0.62\columnwidth]{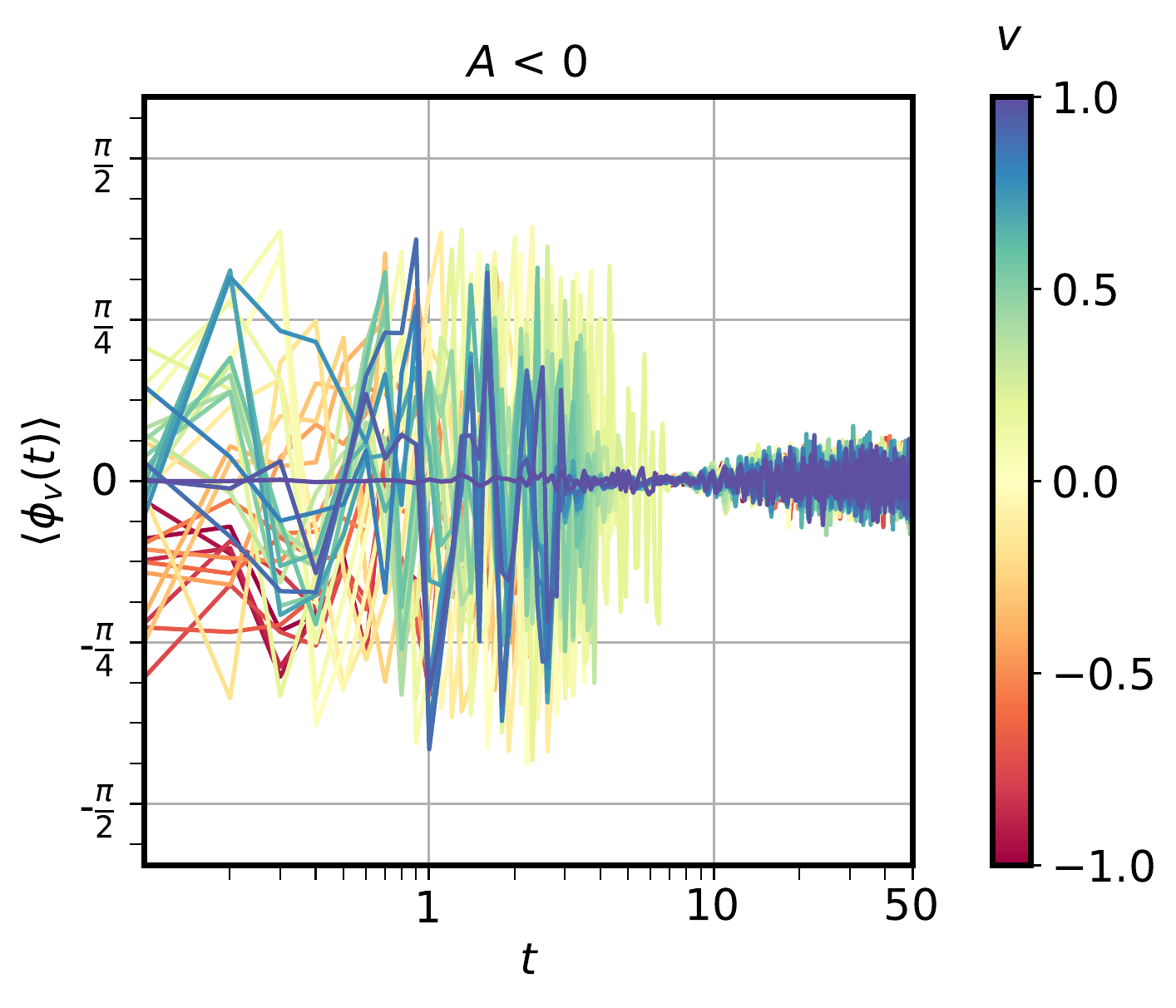}
	\hspace{0.3cm}\includegraphics[width=0.62\columnwidth]{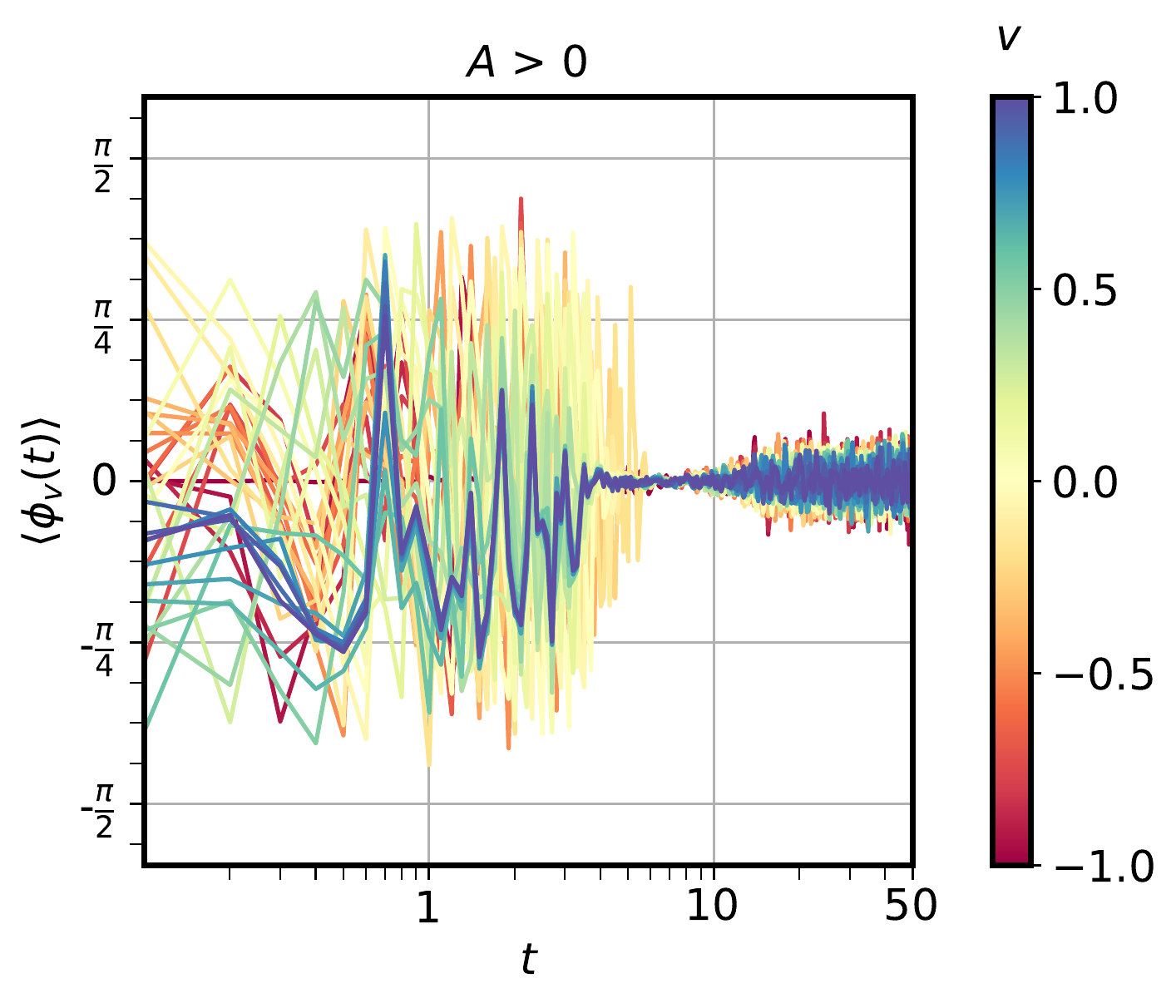}\\
	\hspace{0.2cm}\includegraphics[width=0.64\columnwidth]{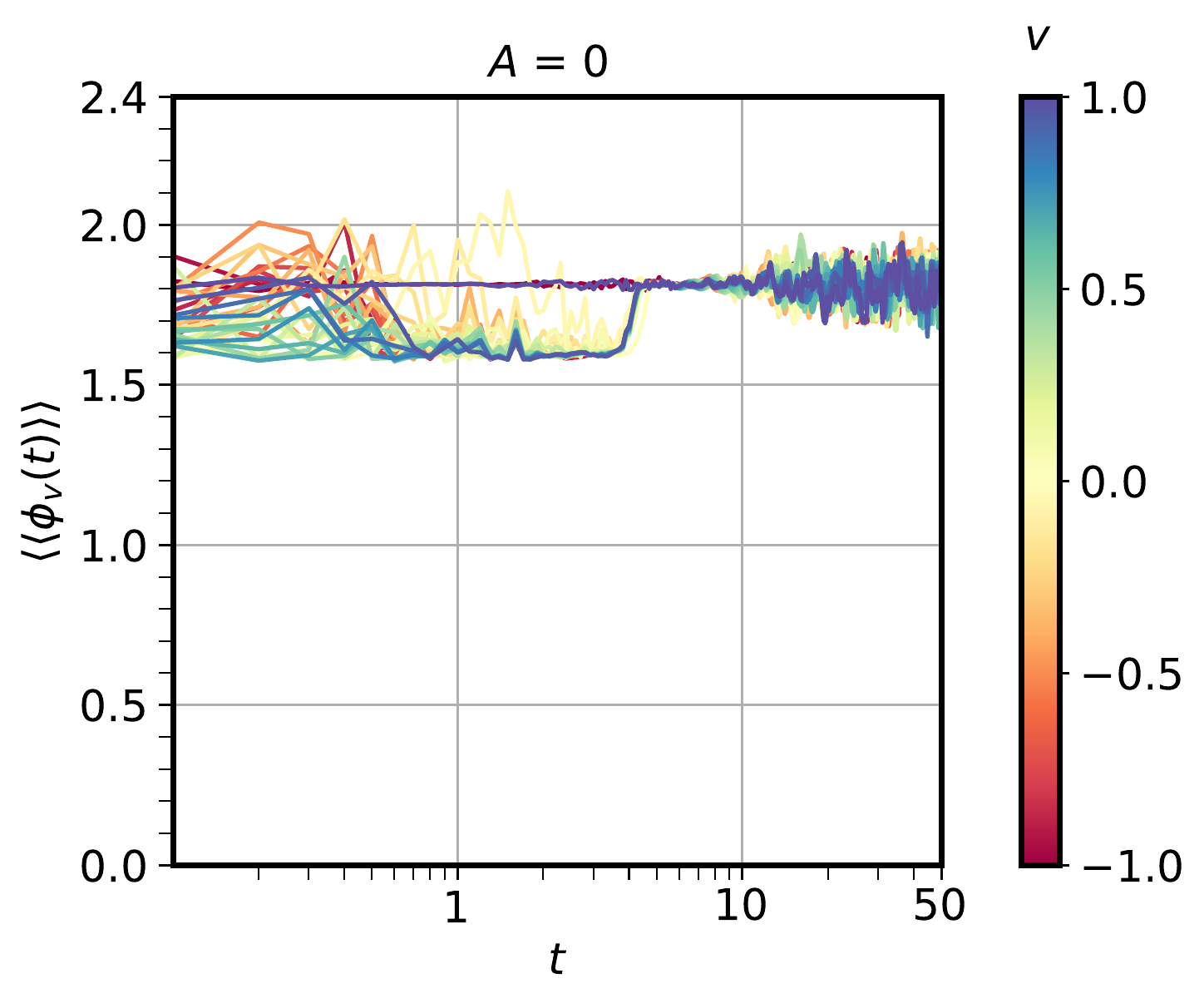}
	\hspace{0.2cm}\includegraphics[width=0.64\columnwidth]{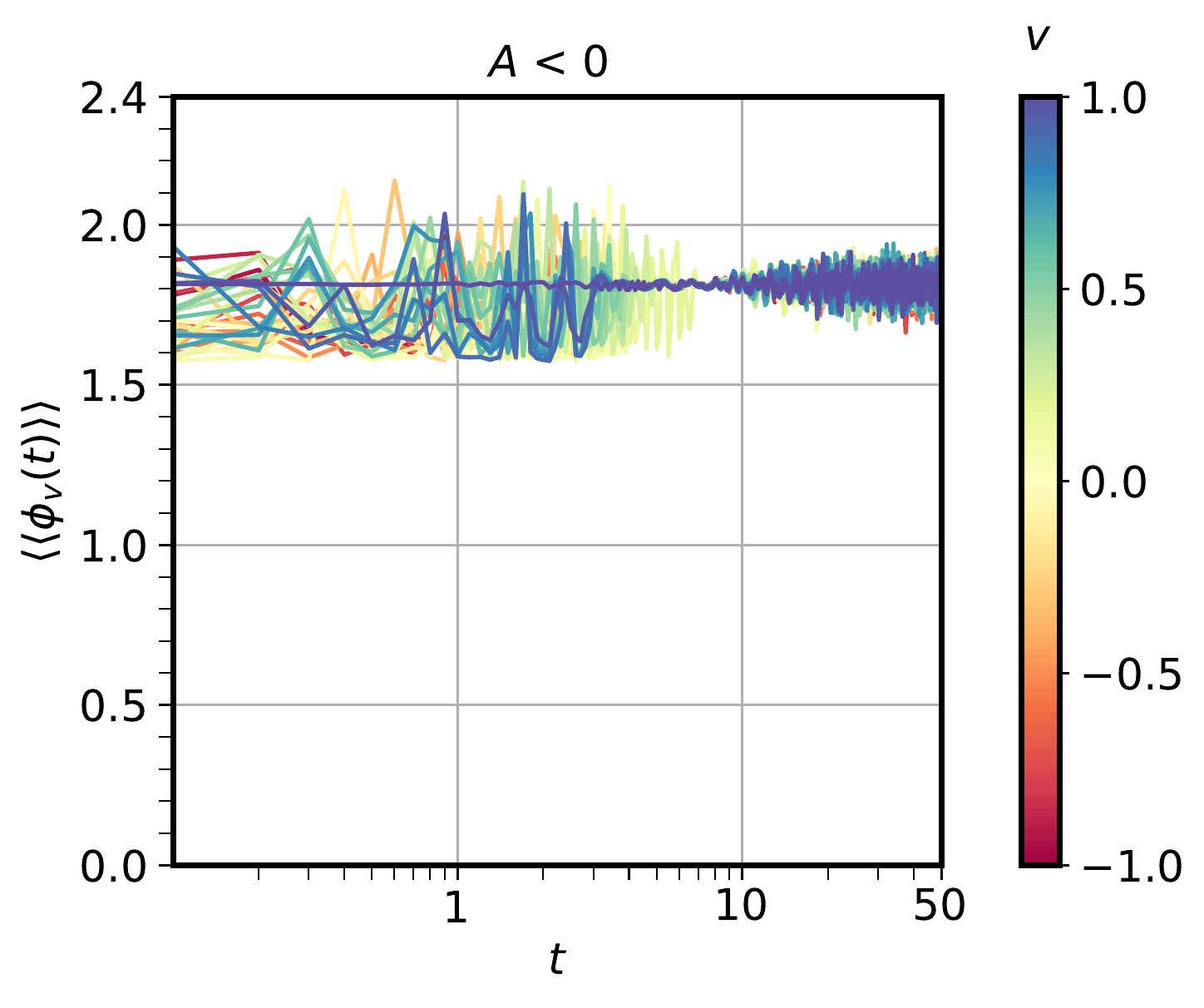}
	\hspace{0.2cm}\includegraphics[width=0.64\columnwidth]{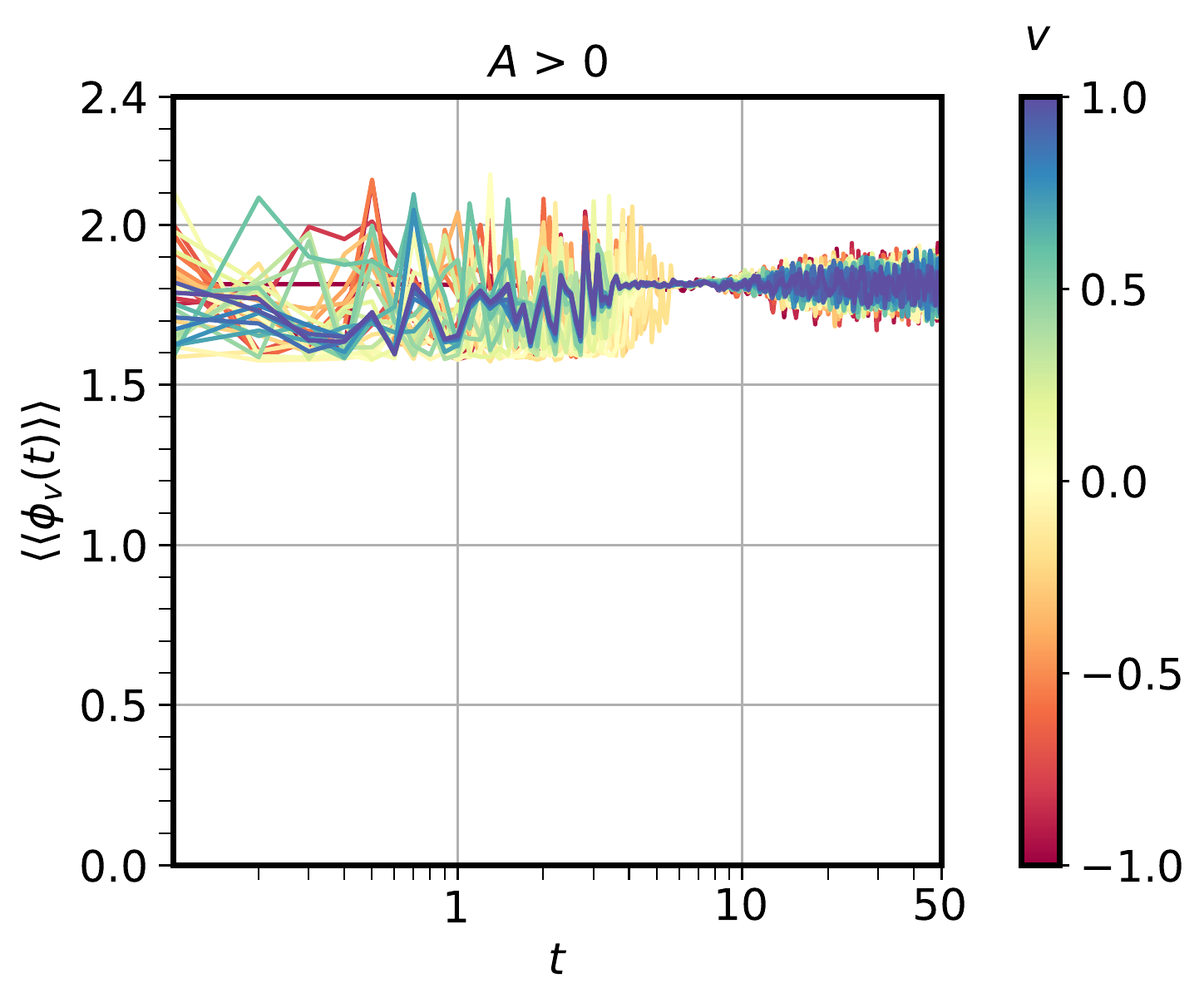}\\
	\includegraphics[width=0.66\columnwidth]{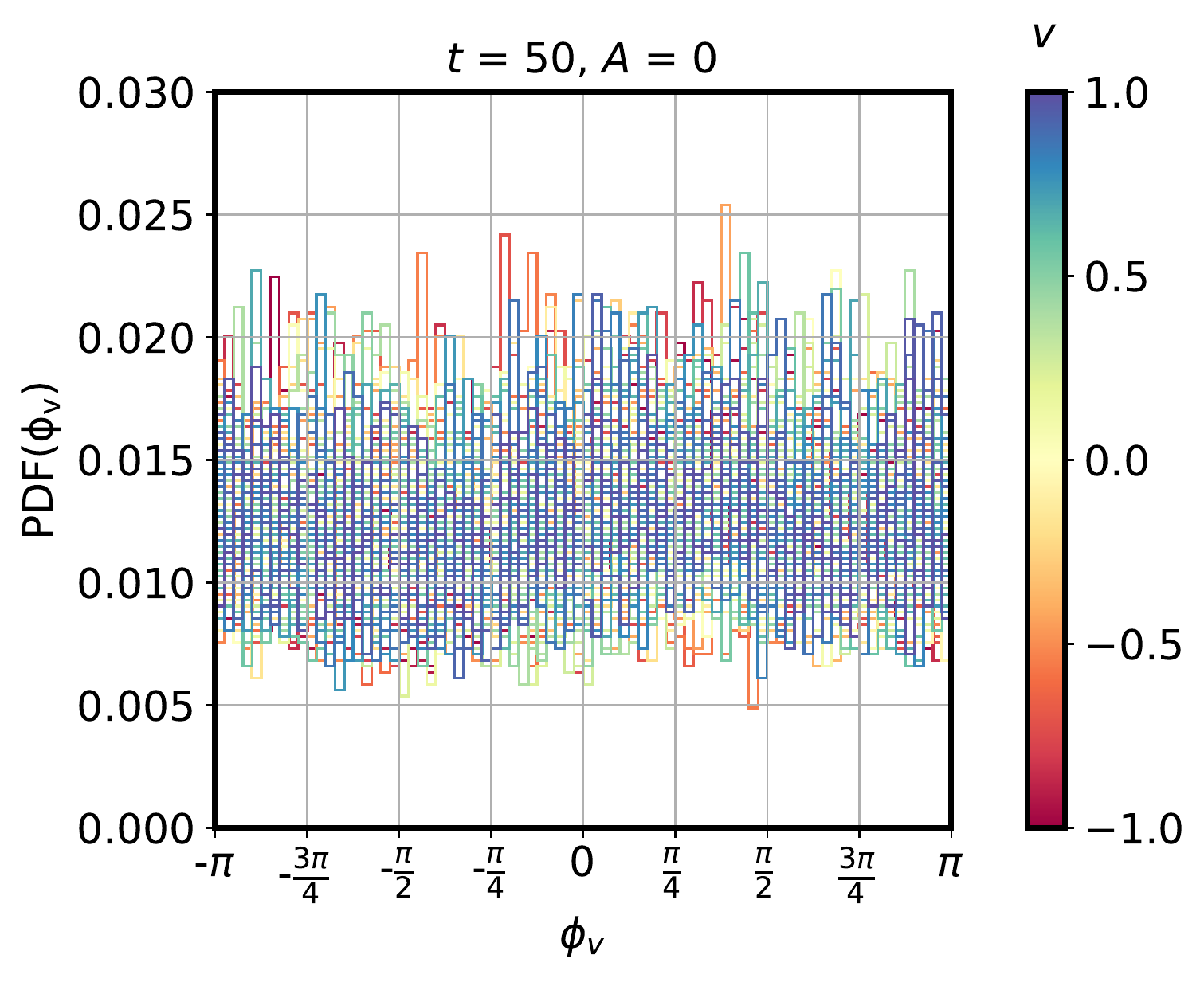}
	\includegraphics[width=0.66\columnwidth]{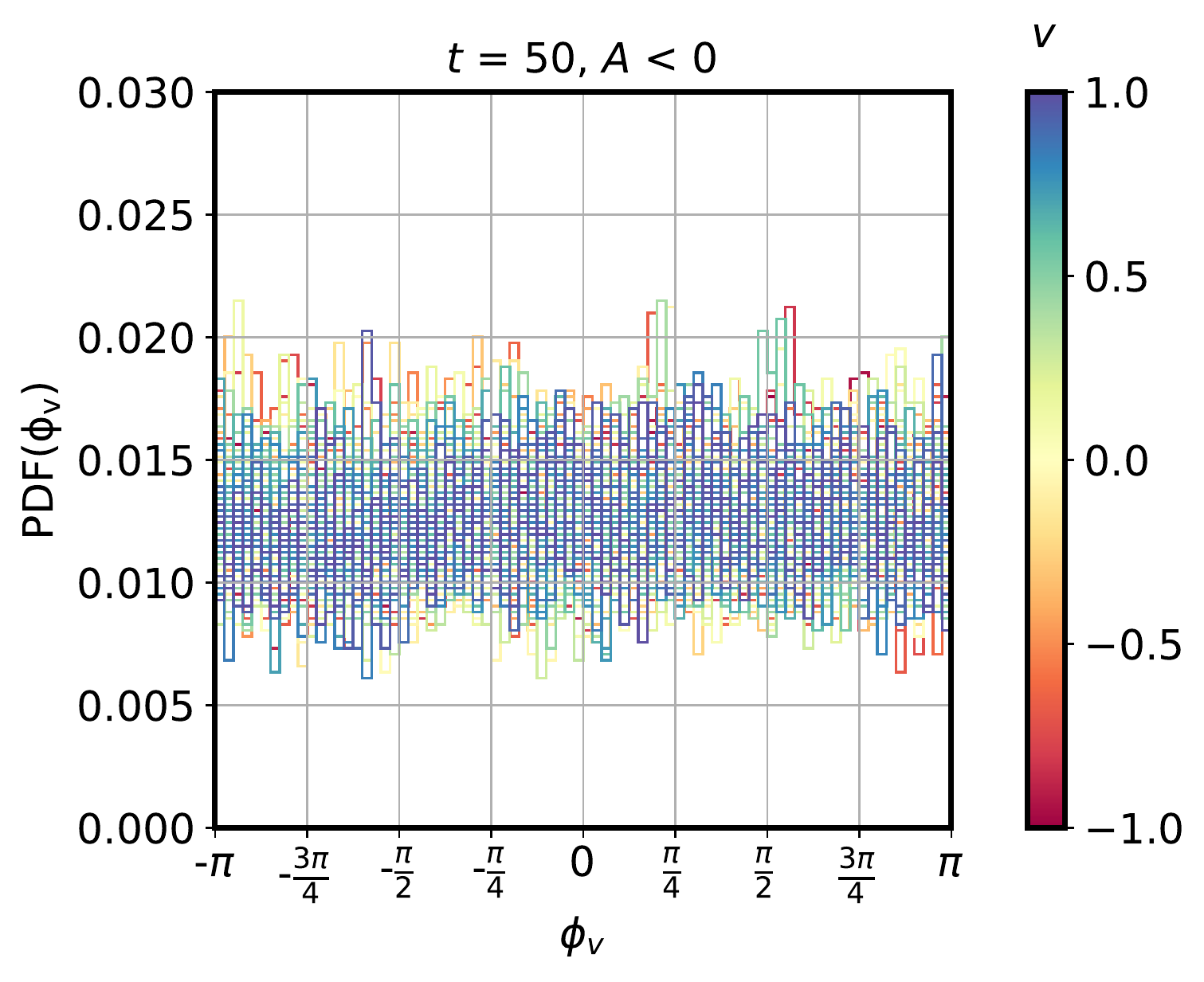}
	\includegraphics[width=0.66\columnwidth]{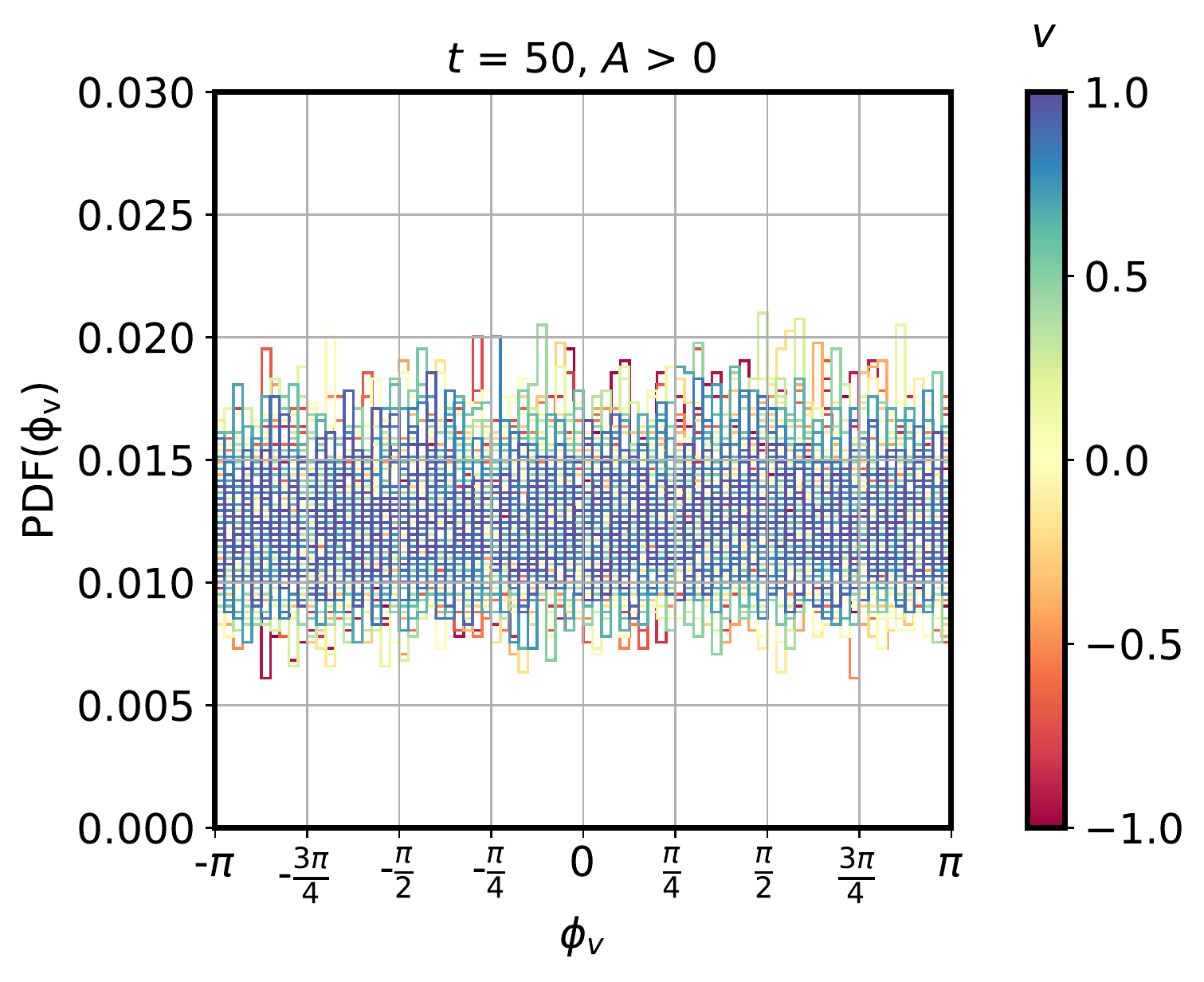}
	\caption{Top: Average of the spatial distribution of $\phi_{v}(x, t)$ at time $t$ = 50. Middle: Standard deviations.  Bottom: Histograms of the spatial distribution of $\phi_{v}(x, t)$. In each plot different velocity modes depicted by color-coding as per color-bar. Left panels show the case $A=0$, middle panels $A<0$, and right panel $A>0$. Note the approximate randomization of $\phi_{v}(x, t)$.}   
	\label{fig6}
\end{figure*}
\begin{figure*}[]
	\includegraphics[width=0.6\columnwidth]{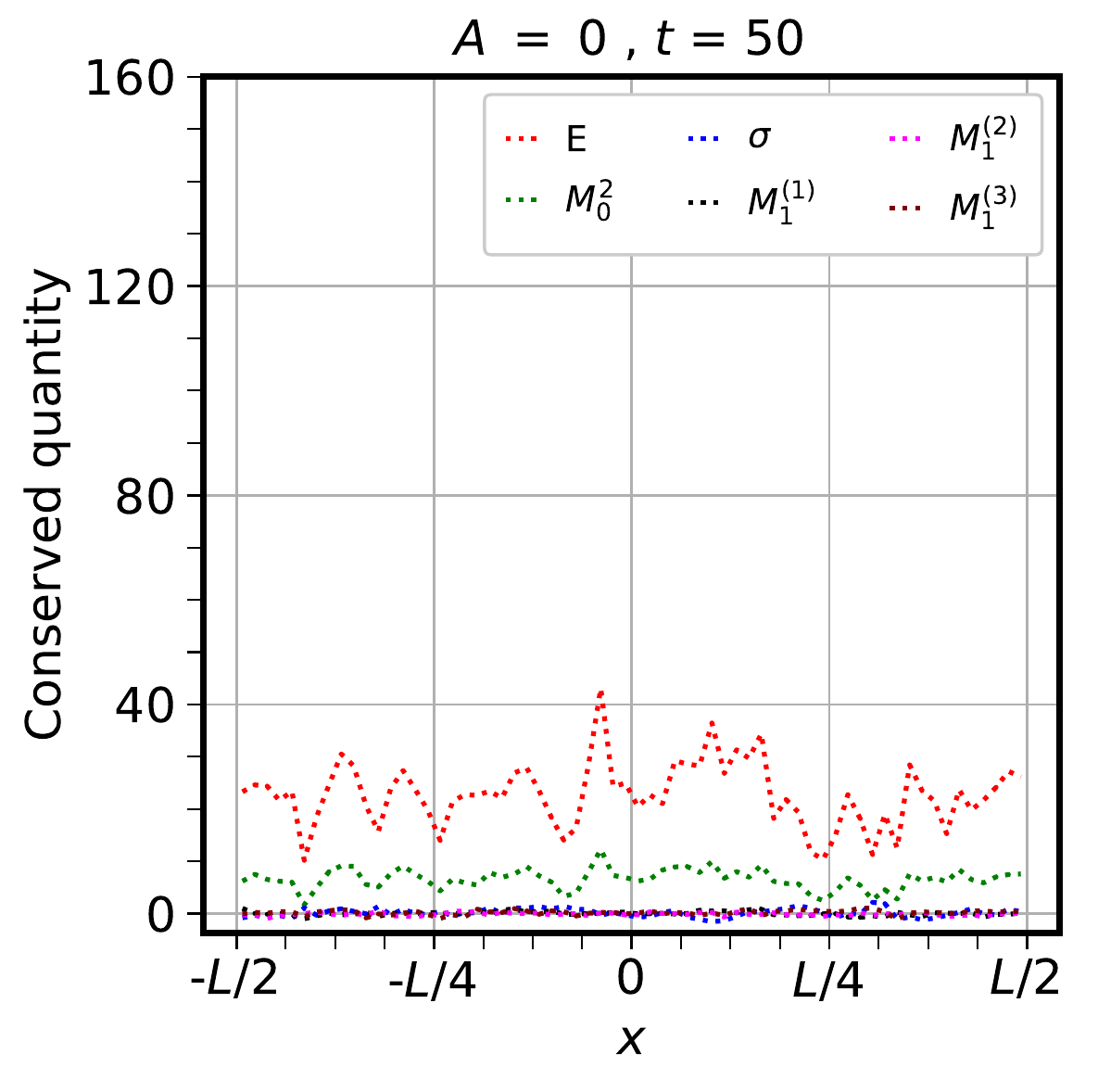}
	\hspace{0.1cm}\includegraphics[width=0.6\columnwidth]{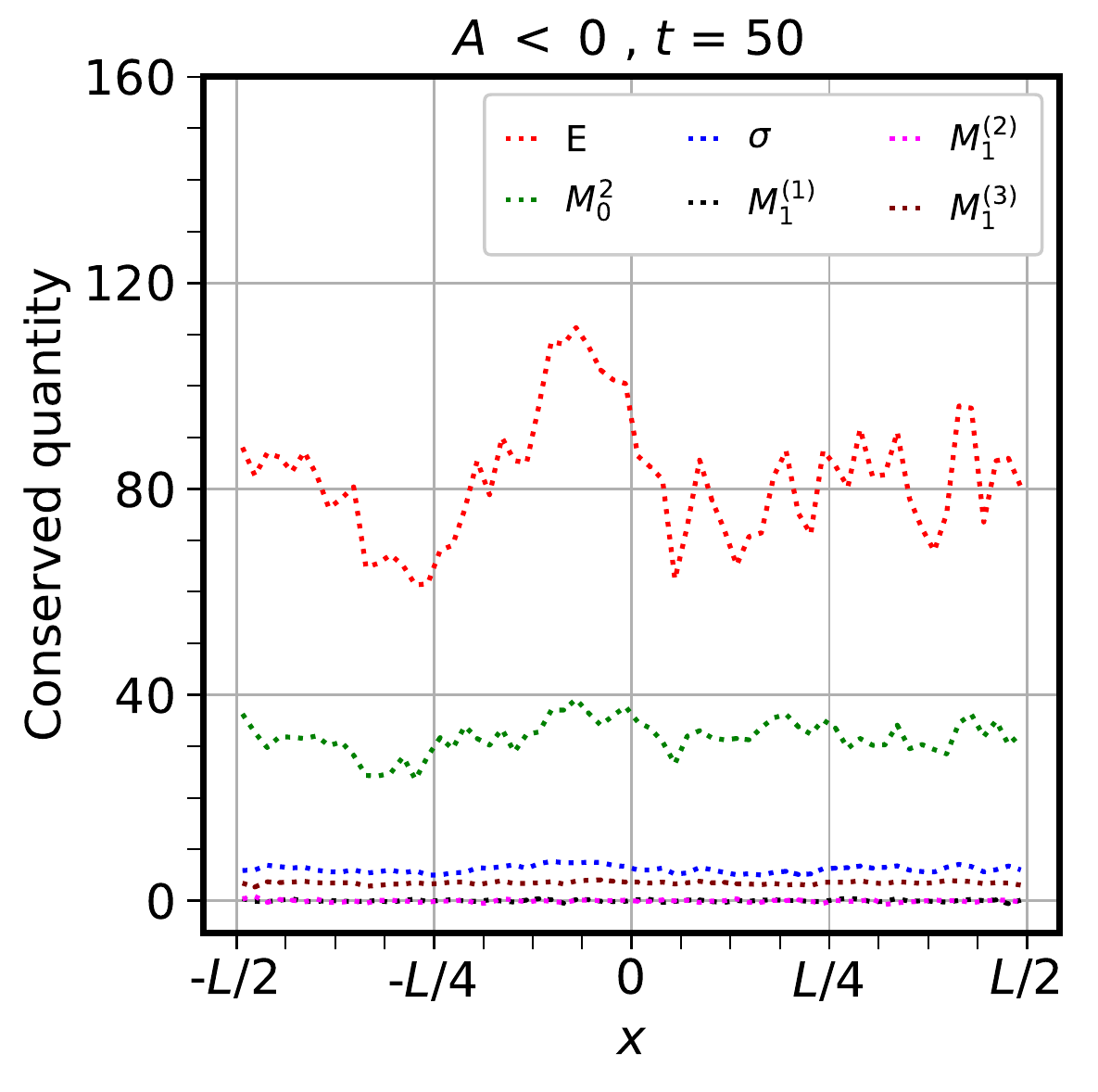}
	\hspace{0.1cm}\includegraphics[width=0.6\columnwidth]{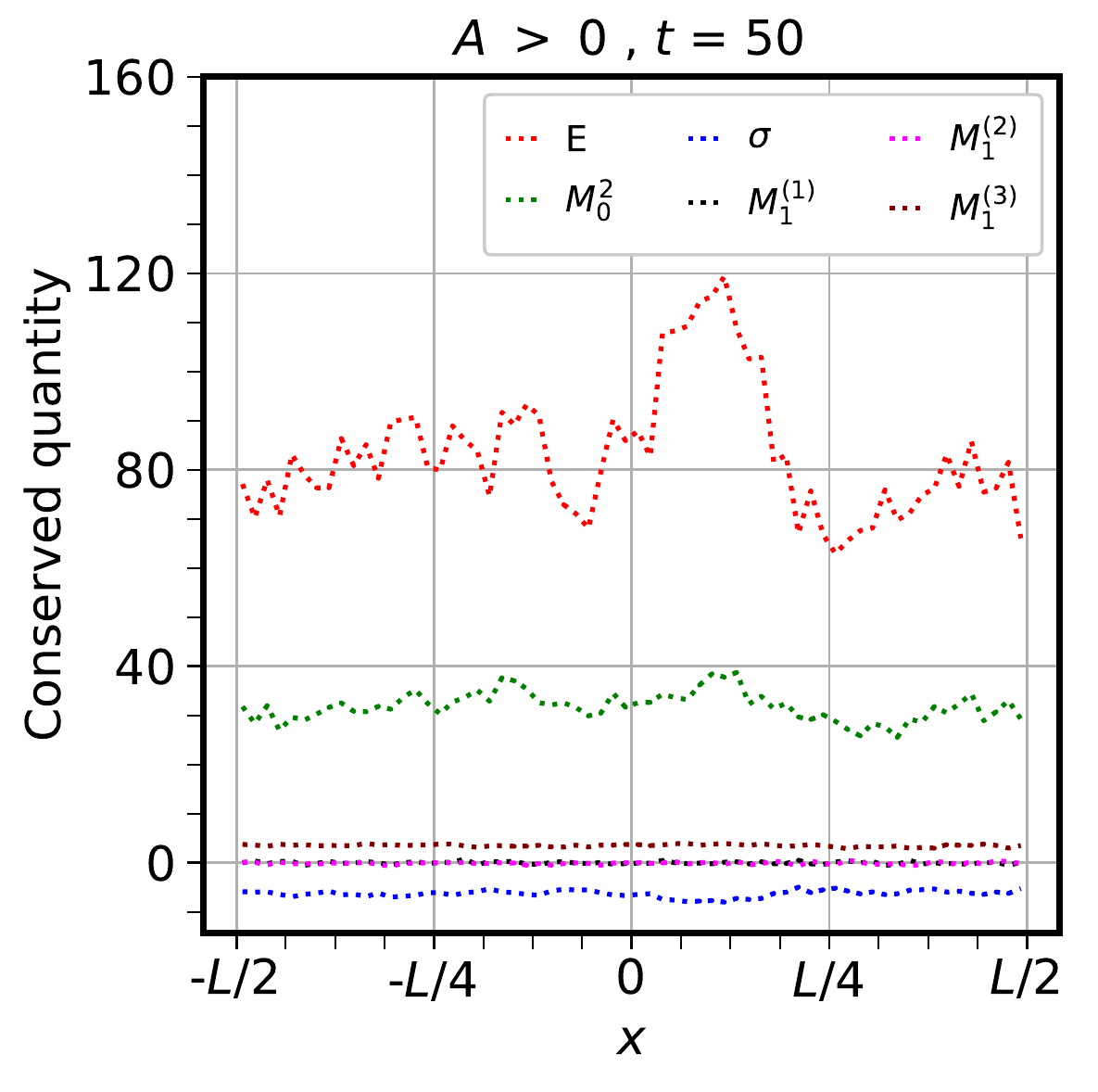}
	\caption{Top: Spatial variations of the conserved quantities ($\vec{M}_{1}, M_{0}^{2}, \sigma, E$) that are supposed to be spatially constant in the extreme nonlinear regime. Left panels show the case $A=0$, middle panels $A<0$, and right panel $A>0$. We believe that the approximate-ness of the constancy of these conserved quantities is related to the approximate-ness of the steady state.} 
	\label{fig7}
\end{figure*}

\begin{figure*}[]
	\includegraphics[width=0.6\columnwidth]{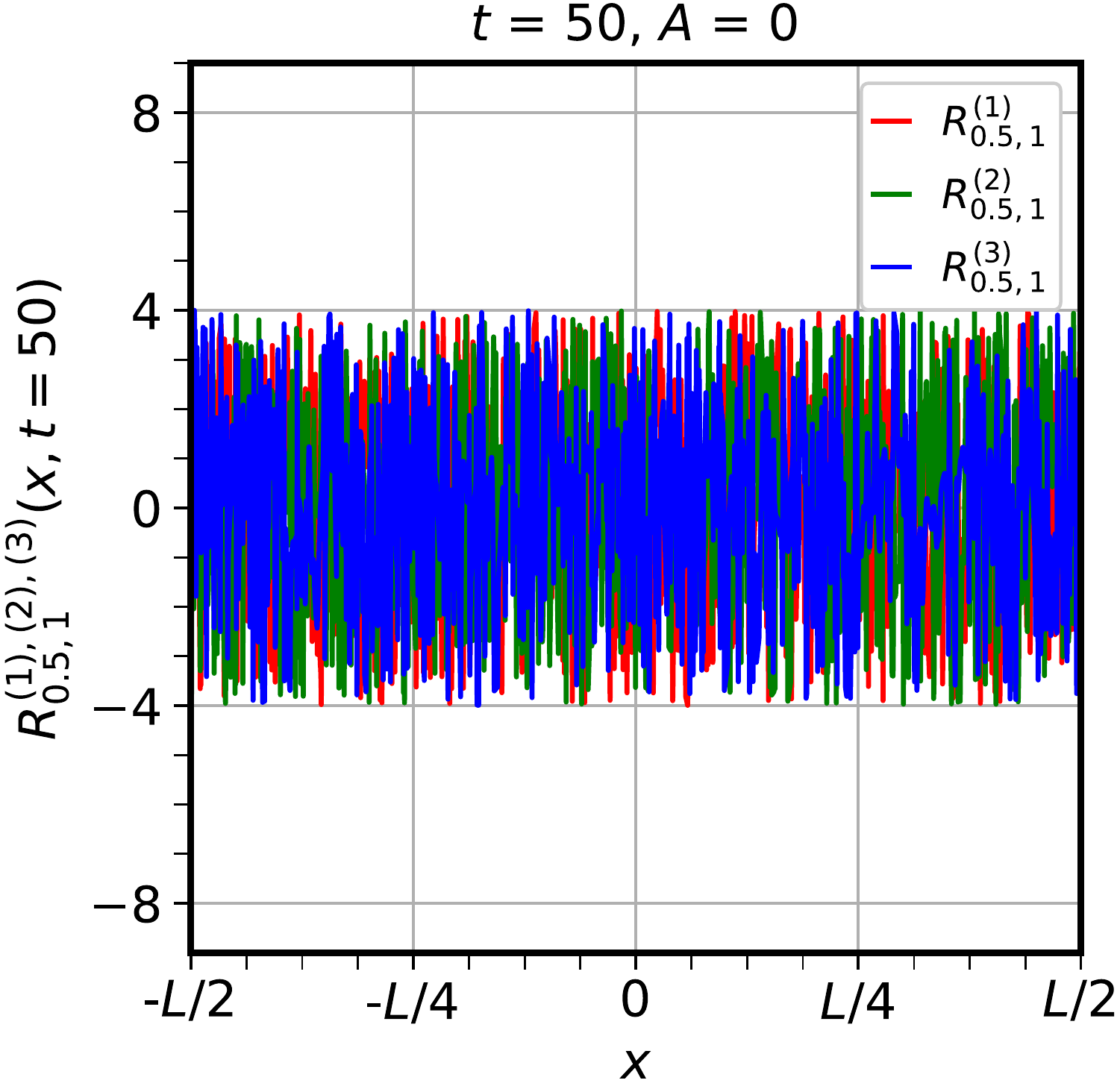}~
	\includegraphics[width=0.6\columnwidth]{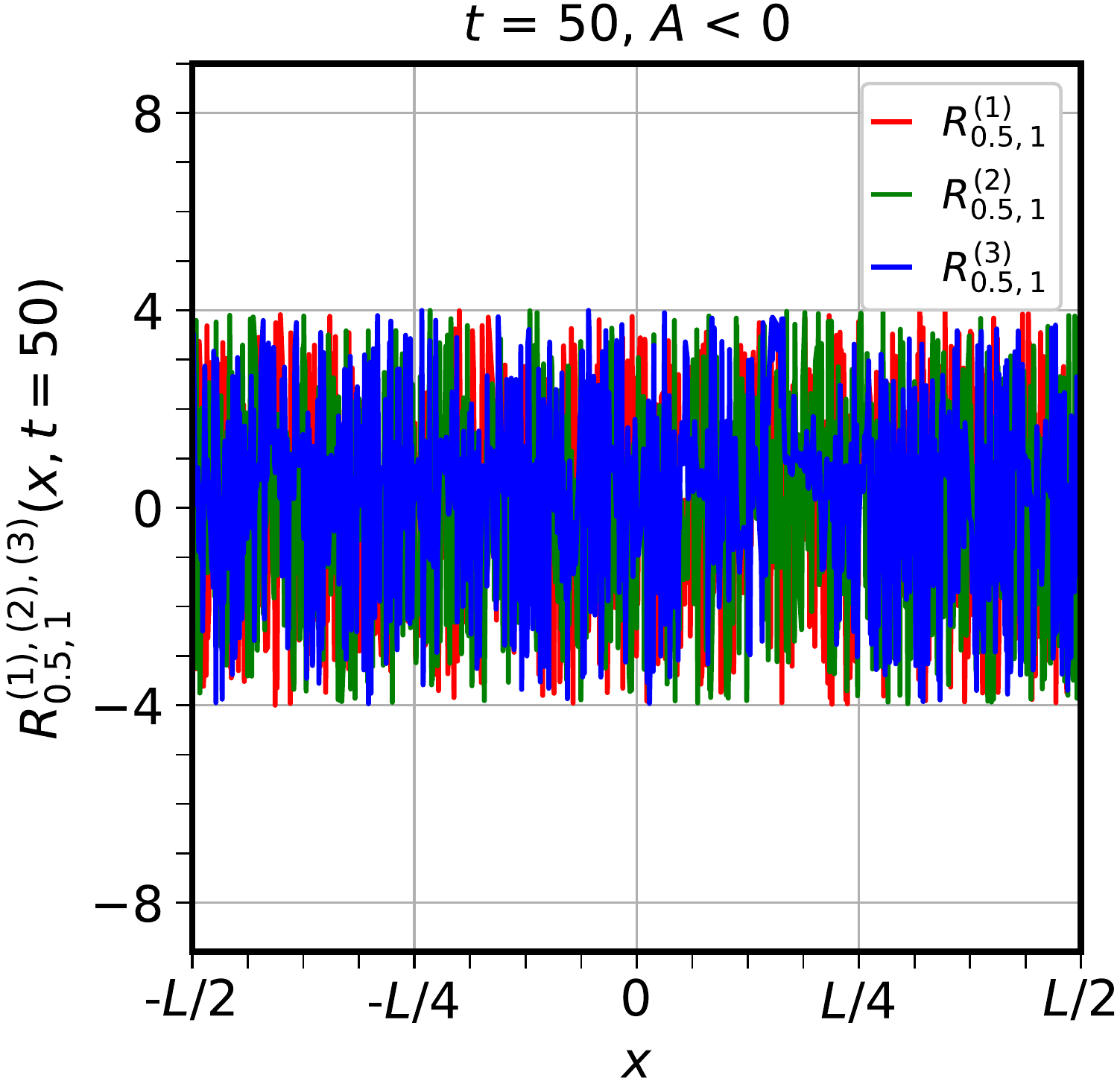}~
	\includegraphics[width=0.6\columnwidth]{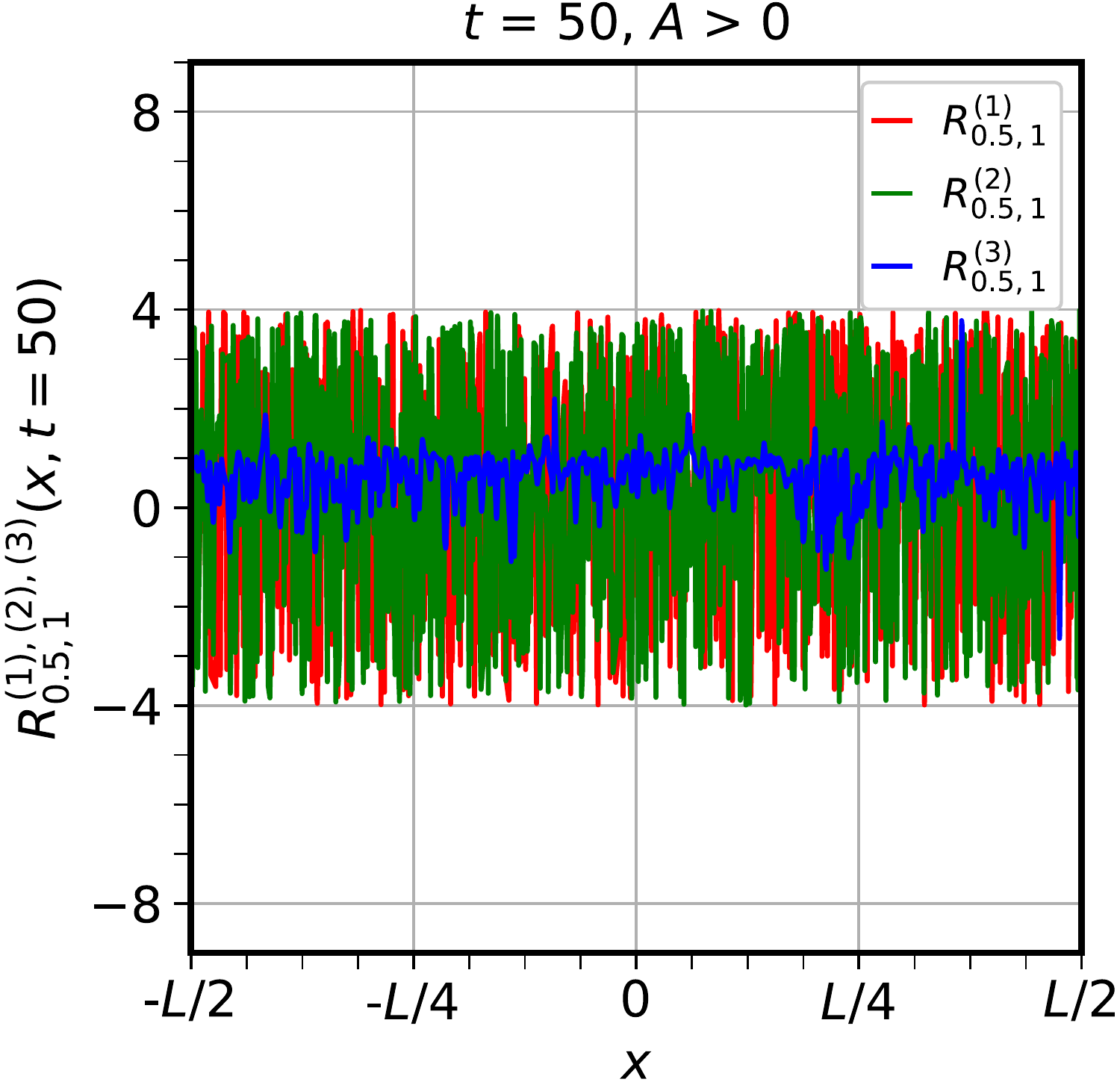}
	\caption{Variation of $R_{v_{1}, v_{2}}^{(i)} (x, t)$ as a function of $x$ at $t = 50$ is shown for $v_{1} = 0.5, v_{2} =1$. The large variation as a function of $x$ shows that the late-time solution is not separable in space and velocity. In other words, at late time the different velocity modes do not show an approximately collective (correlated) evolution in space.} 
	\label{fig8}
\end{figure*}

\section{Numerical Study}\label{Num}

In the following subsections, we now discuss our own numerical strategy to solve the equations of motion, the late-time behavior of the solution, and their relation to the analytical claims made in Sec.\,\ref{ana}. 

We developed a code for solving Eq.$\eqref{2}$. In our code, we discretize each of the spatial directions as well as the velocities, considering $N_{x}$ spatial modes and $N_{v}$ velocity modes, to get a total of $3\,N_{x}\,N_{v} $ coupled nonlinear ODEs, where the factor of 3 comes due to the 3 elements of each polarization vector. It solves these coupled nonlinear ODEs as a function of time using {\tt python}'s {\tt zvode} solver, which is a complex-valued variable-coefficient ordinary differential equation solver in {\tt python} which implements Backward Differentiation Formula for doing numerical integration. The spatial derivatives at each spatial point are computed using {\tt python}'s {\tt scipy.fftpack.diff} package, which uses the Fast Fourier Transform method for calculating derivatives.

For the computations shown in this paper, we choose a periodic boundary condition in space such that, for any time $t$ and any velocity mode $v$, our solution satisfies
\begin{equation}\label{40}
\vec{P}_{v}(x, t) = \vec{P}_{v}(x+L, t)\,.
\end{equation}
We choose initial conditions such that all the neutrinos, with any velocity, are emitted as purely electron flavored states at every point in space. To trigger the flavor evolution, a perturbation of $10^{-6}$ is used as an initial seed for the transverse components of the polarization vectors, for every velocity mode at the midpoint of the 1D box. This choice is arbitrary, but motivated by trying to seed all wavelength modes of the instability equally, as opposed to using, say, homogeneous initial conditions that favor a specific mode. We ensure that the size of the box $L$ and the maximum time $t_{\rm{final}}$ up to which we solve the equations, are such that $2\,t_{\rm{final}} < L$, as a result of which the mode with the largest velocity does not see the boundary of the box when emitted from the centre of the box, where we seed the instability. 

Specifically, we choose a 1D box of size $L = 115$  and discretize it into $2^{12}$ spatial bins to solve Eq.\eqref{2} numerically  upto a time of $t_{\rm{final}} = 50$. We discretize the ELN distributions into $2^{7}$ velocity modes, giving rise to a total of $3\times 2^{12} \times 2^{7} = 1572864$ coupled ODEs. These choices are optimized to obtain sufficient accuracy and precision, as we show in Appendix\,\ref{sec:appendix}. To show the dependence of our results on the ELN, we consider three qualitatively different continuous ELN distributions shown in Fig.\,\ref{fig1}. We choose these ELNs in such a way that they all have a zero crossing in their velocity distribution for fast flavor conversion to occur, but they have lepton asymmetries that are either zero, or negative, or positive. Fig.\,\ref{fig2} clearly indicates that the system becomes approximately steady already at $t_{\rm{NL}} \approx 10$, and to ensure that we were deeply in the nonlinear regime, we ran our code until $t=50$. 

We define some notation for convenience:  
\begin{align}
\vec{S}_{v}^{\perp}(x, t) &= s_{v}^{(1)}(x, t)-i s_{v}^{(2)}(x, t)\nonumber\\ 
&= \abs{ \vec{S}_{v}^{\perp}(x, t)}e^{i \phi_{v}(x, t)}\,\label{41}, 
\end{align}
where $\phi_{v}(x, t)$ encodes the correlation between the two transverse components of the polarization vector for each velocity mode, with $\abs{ \vec{S}_{v}^{\perp}(x, t)}$ describing the length of the vector in the ${\hat{\bf e}_1}-{\hat{\bf e}_2}$ plane. We will denote all the spatially averaged quantities over the whole box of size $L$ as $\langle\cdots\rangle$, the velocity averaged quantities by $\overline{\cdots}$, the probability distribution for the spatial distribution of some quantity as $\rm{PDF}(\cdots)$ and the standard deviation for that distribution as $\langle \langle\cdots\rangle\rangle$.

\subsection{Late Time Behavior of Parallel and Perpendicular Components}
In Fig.\,\ref{fig3}, we show the time evolution of the spatial average and standard deviation of $s_{v}^{(3)}(x, t)$,  for all three types of lepton asymmetry. In Fig.\,\ref{fig4}, we show the histograms of $s_{v}^{(3)}(x, t)$ and $|\vec{S}_{v}^{\perp}(x, t)|$, at a time $t=50$ and for various velocity modes color-coded by their velocity. We have checked that these histograms are themselves quasi-stationary at late time. These results indicate approximate stationarity in time in the nonlinear regime for all velocities. Note how $\bigl\langle s^{(3)}(t)\bigr\rangle$ separates into two cohorts, dictated essentially by the sign of $v$, at late time in the cases $A>0$ and $A<0$. This behavior is very similar to spectral swaps. As explained in Sec.\,\ref{ana}, which modes remain close to their initial state and which move away depends on $A$, $B$, and the higher moments.

The spatial distribution of $s_{v}^{(3)}$, i.e., ${\rm PDF}(s_{v}^{(3)})$, and the distribution of $|\vec{S}_{v}^{\perp}|$, i.e., ${\rm PDF}(|\vec{S}_{v}^{\perp}|)$, shown in Fig.\,\ref{fig4}, are obviously related because $\abs{\vec{S}_v} = 1$. The latter is obtained from the former by multiplying with the Jacobian of the transformation from $s_{v}^{(3)}$ to $|\vec{S}_{v}^{\perp}|$:
\begin{equation}\label{42}
\begin{aligned}
{\rm PDF}(|\vec{S}_{v}^{\perp}|) & = {\frac{\abs{\vec{S}_{v}^{\perp}}}{\sqrt{1-\abs{\vec{S}_{v}^{\perp}}^{2}}}}\,{\rm PDF}(s_{v}^{(3)})\,.
\end{aligned}
\end{equation}
Consider the case with $A = 0$. In this case, ${\rm PDF}(s_{v}^{(3)})$ is approximately uniform (see top left panel of Fig.\,\ref{fig4}) and as a result, ${\rm PDF}(|\vec{S}_{v}^{z_{\perp}}|)$ should be a sharply peaked distribution for every $v$ with a peak very close to $|\vec{S}_{v}^{\perp}| = 1$ and a tail at $|\vec{S}_{v}^{\perp}| = 0$. In bottom left panel of Fig.\,\ref{fig4}, we recover this unsurprising feature. For cases with $A < 0$ ($A > 0$), the negative (positive) velocity modes show a peaked distribution in $s_{v}^{(3)}$ almost close to 1, implying a more broadened distribution for $|\vec{S}_{v}^{\perp}|$ between 0 to 1, as seen in the middle and right panels.

Top panels of Fig.\,\ref{fig5} show the oscillatory nature of the spatial variation of $s_{v}^{(3)}(x, t)$,  at time $t = 50$ when the system becomes fully nonlinear. The bottom panels of Fig.\,\ref{fig5}, show that the spatially averaged polarization vectors are depolarized for $A = 0$, with $\langle s_{v} \rangle \approx 0$, but not so in case of nonzero lepton asymmetry. For $A\neq0$, the polarization vectors have not flipped close to $v\approx-1$ ($v\approx+1$) for $A<0$ ($A>0$), as predicted by our analysis of fast spectral swaps. One interesting point to note from the bottom-middle and bottom-right panels of Fig.\,\ref{fig5}, with spectra $A \neq 0$, is that the functional behavior in the velocity space shows a mirror symmetry when we switch the sign of $A$. This is not surprising, because our chosen ELNs have a symmetry $G_{A<0}(v)=-G_{A>0}(-v)$.

\subsection{Behavior of the Phase $\phi_{v}(x, t)$}
We obtain the phase $\phi_{v}(x, t)$, for a given space-time point $(x, t)$ with a velocity $v$, in the range $[-\pi, \pi]$ using the python package {\tt math.atan2} which takes $s_{v}^{(1)}(x, t)$ and $s_{v}^{(2)}(x, t)$ as inputs, obtained from our numerical simulation, and returns the principal value of ${\rm arg}\big({\bf S}^\perp_v(x,t)\big)$ as the output.

Fig.\,\ref{fig6} shows that the phase is almost uniformly random, and its mean and variance are consistent with that expected of a uniformly random distribution over space, for every velocity mode and for any type of lepton asymmetry. Initially, for each velocity mode the transverse components of a polarization vector $\vec{S}_{v}(x, t)$ are correlated, but in the extreme nonlinear regime such a kinematic phase coherence is lost.  

\subsection{Conserved Quantities}
In Sec.\,\ref{ana} we showed that $\vec{M}_{1}$ is almost constant in space in the steady state approximation, which allowed us to go in a rotating frame where $\widetilde{\vec{M}}_{0}$ has a motion equivalent to a gyroscopic pendulum. Further, the pendulum's length $M_{0}$, spin $\sigma$, and energy $E$ can be seen to be spatially conserved. In this subsection, we undertake a numerical survey to verify if these quantities indeed satisfy the above analytical claim. 

To calculate the different multipole moments numerically we used the late-time solution for $\vec{P}_{v}(x, t)$ from our code and integrated over all the velocity modes from -1 to 1 at each spatial point with an appropriate weightage factor, corresponding to the Legendre polynomial of a degree equivalent to the multipole number of that particular moment. One technical issue is that the quantities $\sigma$ and $E$ involve infinite series in the multipole moments which are numerically impossible to calculate. Thankfully, the spatially averaged components of the polarization vectors, shown in the bottom panel of Fig.\,\ref{fig5}, behave very smoothly as a function of velocity and can be adequately approximated by the first few multipole moments. 

Fig.\,\ref{fig7} shows the numerical results for the spatial variation of the conserved quantities at late time. Most of the conserved quantities are approximately constant over space, but the energy $E$ of the gyroscopic pendulum does not appear to be sufficiently constant (relative to, say, the variation of individual polarization vectors). Even at late time the stationarity of the components of $\widetilde{\vec{S}}_{v}(x, t)$ is approximate, which reflects in the degree of variation in the spatial behavior of the conserved quantities. In contrast, if one were to look at the velocity averaged $\overline{\bf P}(x,t)$, that is much noisier even at late times, not unlike what is shown in the upper panels of Fig.\,\ref{fig5}. 

To verify if Eq.\eqref{2} has a  separable solution for the components of $\vec{S}_{v}(x, t)$ in position and velocity space we define a ratio as $R_{v_{1}, v_{2}}^{(i)} (x, t) = \frac{s_{v_{1}}^{(i)}(x, t)}{s_{v_{2}}^{(i)}(x, t)}$ for the $i^{\rm{th}}$ component of the polarization vector at a time $t$ in the full nonlinear regime for two different velocity modes $v_{1}, v_{2}$. This will be spatially constant if the solution is separable in $x$ and $v$. To calculate this quantity we considered the solution at $t=50$ with the modes $v_{1} = 0.5$ and $v_{2} = 1$ for all the three components ($i = 1, 2, 3$) of the polarization vector, and plotted $R_{v_{1}, v_{2}}^{(i)} (x, t)$ as a function of $x$. Fig.\,\ref{fig8} shows a large deviation from a constant, confirming a non-separable solution in the nonlinear regime.

\section{Conclusions}\label{Con}

Our goal in this paper was to study the late-time behavior of fast oscillations. To this end, we considered a model in 1+1D and gave an analytical as well as a numerical understanding of its flavor dynamics. Here we summarize our main results:

\begin{itemize}

\item The system reaches an approximately steady state in time in the extreme nonlinear regime. 
\item In a rotating frame the spatial evolution of all the polarization vectors is a velocity-dependent precession about a common axis.
\item This common axis acts as a gyroscopic pendulum with a fixed length, spin, and energy leading to an oscillatory behavior in position space.
\item The steady state solution for the components of the polarization vector is not separable in position and momentum space. 

\item The phase of the transverse component of the polarization vector at different spatial locations becomes randomly distributed over the interval $[-\pi, \pi]$ at late time, for all velocity modes and with any value of lepton asymmetry.

\begin{figure*}[!t]
	\includegraphics[width=0.8\columnwidth]{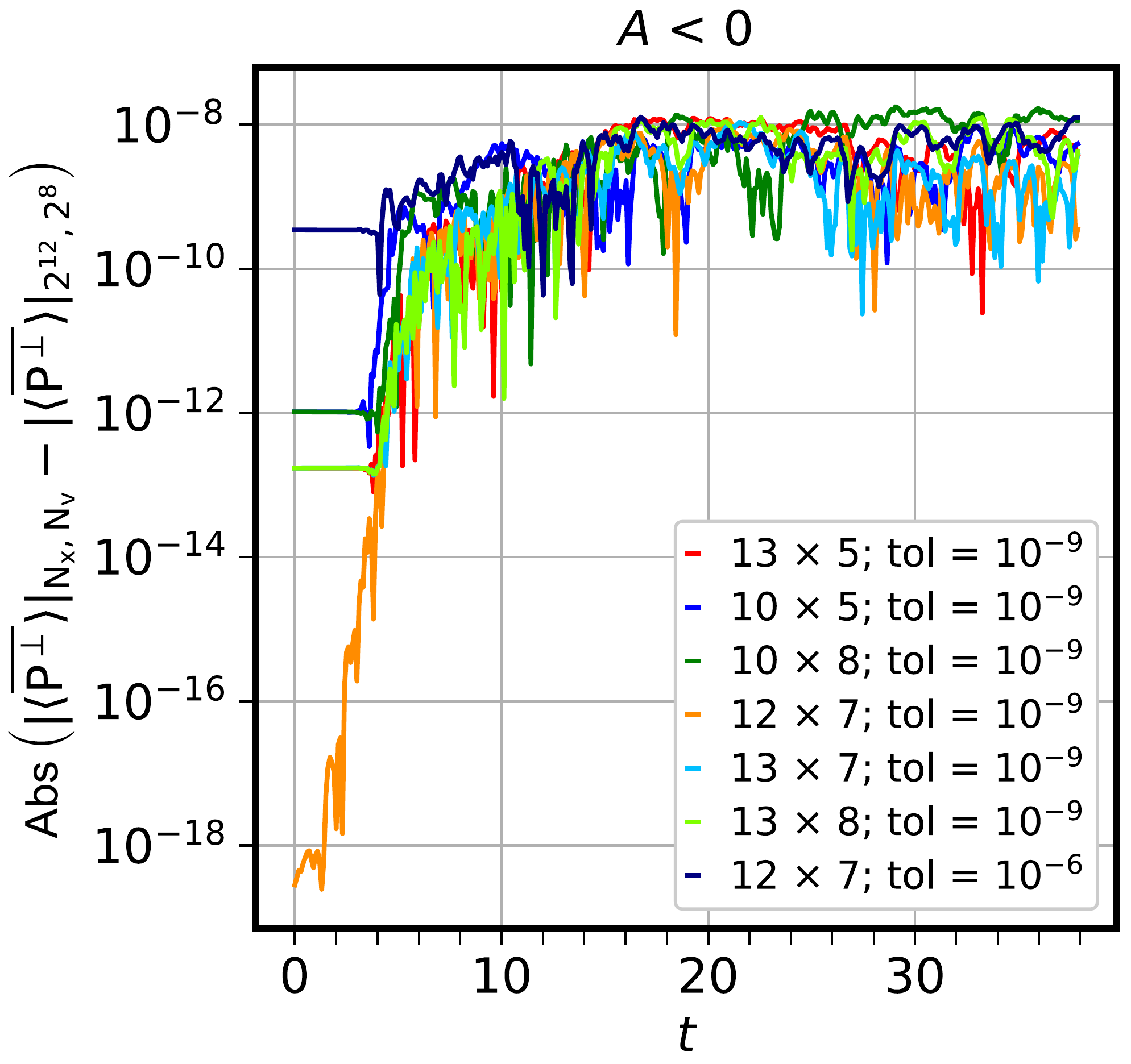}
	\hspace{2cm}\includegraphics[width=0.79\columnwidth]{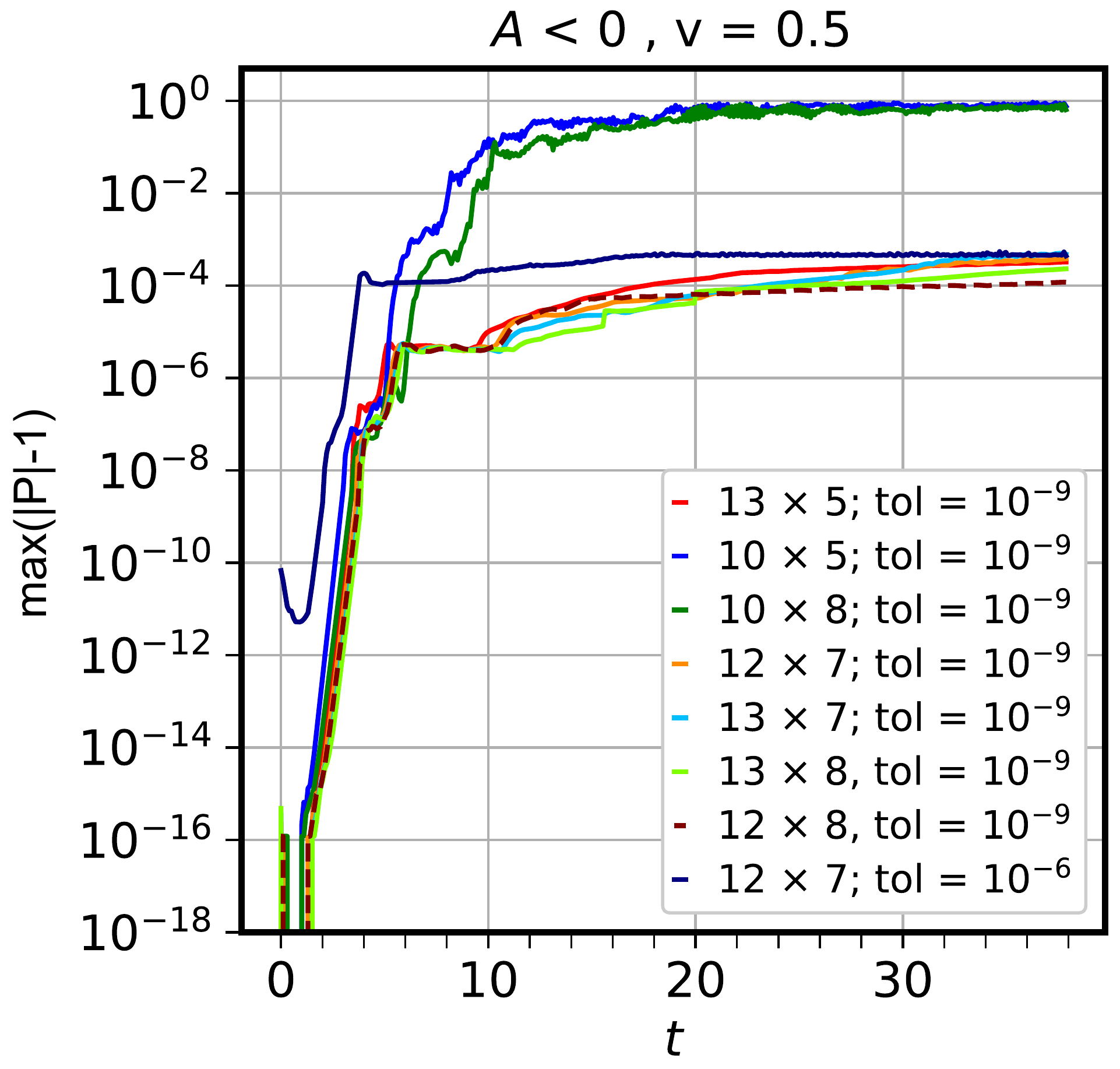}
	\caption{Left: Precision of our calculation, estimated by convergence of the computed $\lvert\overline{\langle{\bf P}^\perp\rangle}\rvert$ for different discretizations $N_{x} \times N_{v}$ (to be read as the exponents of 2, i.e., $12\times7$ refers to $N_x=2^{12}$ and $N_v=2^{12}$), as a function of time $t$, shown for the $A < 0$ case. Right: Accuracy of our calculation, estimated by maximum departure (over all space) of the magnitude of ${\bf P}_{v=0.5}$ from unity, as a function of time $t$, shown for the $A < 0$ case and different discretizations. Absolute and relative tolerances were taken to be same and mentioned alongside.}
	\label{fig9}
\end{figure*}

\item The velocity dependence of the components of the polarization vector is controlled by the lepton asymmetry of the system. Systems with zero lepton asymmetry have a ``decoherent'' behavior for all velocity modes, with $\langle s_{v}^{(1)} \rangle = \langle s_{v}^{(2)} \rangle = \langle s_{v}^{(3)} \rangle = 0$ for every $v$. In case of nonzero lepton asymmetry, the decoherence is not complete, but $\langle s_{v}^{(1)} \rangle$ and $\langle s_{v}^{(2)} \rangle$ still remain zero for every velocity mode, while $\langle s_{v}^{(3)} \rangle$ show some dependence on velocity to conserve the lepton asymmetry. 

\item The behavior of $\langle s_{v}^{(3)} \rangle$ in the velocity space shows a mirror symmetry if the sign of the lepton asymmetry is flipped. 

\item $\langle s_{v}^{(3)} \rangle$ change with time in a way reminiscent of spectral swaps. At late time, the configuration becomes steady and is given by a velocity dependent ``swap'' function. The constraints on $A$ and $B$ must be obeyed for such a ``swap''.

\end{itemize}

We hope that these results provide some insight into the late-time flavor dynamics associated with fast flavor conversions of self-interacting neutrinos. Although we have chosen to study a simple system in 1+1D, some of these physics results may be useful to understand the more realistic scenario associated with neutrinos in a core-collapse supernova.   


\appendix
\section{Error Estimate}
\label{sec:appendix}

In Fig.\,\ref{fig9} (left panel), we illustrate the precision to be expected of our calculation. We solved the problem, outlined in the main text, for a sequence of increasingly finer discretizations for  space ($N_x$) and velocity ($N_v$), with the $\log_2$ of the numbers of divisions noted in the legend of the figure. We checked for convergence by comparing the results for these discretizations with the discretization $N_x=2^{12}$ and $N_v=2^8$. Our results indicate that a discretization of $N_x=2^{12}$ and $N_v=2^{7}$ is at most ${\cal O}(10^{-8})$ off from yet finer discretizations. 

In Fig.\,\ref{fig9} (right panel), we show the  accuracy expected of our calculation. We check if the length of the polarization vectors remain fixed at 1. The error we incur on this is a lower bound on the error in our calculations. We find that our chosen discretization, $N_x=2^{12}$ and $N_v=2^{7}$,  does as well as finer discretizations, making an error of ${\cal O}(10^{-4})$ at late time. Considering that the quantities we are interested in are close to ${\cal O}(1)$, this error is tolerable. Also, we find that the tolerance dictates how fast the errors increase initially, but the errors plateau out at roughly $\sim100$ times the tolerance. Whereas, if one is interested in a solution that is accurate to ${\cal O}(10^{-4})$, it is less expensive to raise the tolerance to $10^{-6}$, without significant penalty on accuracy and with significant gain in speed.

\vspace{-0.4cm}
\section*{Acknowledgements}

This work is partially supported by the Dept. of Atomic
Energy (Govt.~of~India) research project \mbox{12-R{\textit \&}D-TFR5.02-0200}, the Dept. of Science and Technology (Govt.~of~India) through a Ramanujan Fellowship, and by the Max-Planck-Gesellschaft through a Max Planck Partner Group awarded to B.D. The numerical computations were done on the {Pride} and {Flock} compute clusters in the Department of Theoretical Physics at TIFR Mumbai.


%

\end{document}